\documentclass[journal]{IEEEtran}
\usepackage[noadjust]{cite}
\usepackage[scale=1]{miama}

\makeatletter
\renewcommand{\fnum@figure}{Fig. \thefigure} 
\makeatother
\usepackage{bibentry} 

\usepackage[noadjust]{cite}
\usepackage{filecontents}
\usepackage{balance}
\usepackage{mathrsfs}
\usepackage{amsfonts}
\usepackage{soul}
\setulcolor{yellow} 

\usepackage[pdftex]{graphicx}
\graphicspath{{../pdf/}{../jpeg/}}
\DeclareGraphicsExtensions{.pdf,.jpeg,.png}
\usepackage[cmex10]{amsmath}
\usepackage{mathabx}
\usepackage{optidef}
\usepackage{array}
\usepackage{mdwmath}
\usepackage{mdwtab}
\usepackage{eqparbox}
\usepackage{url}
\usepackage{color,soul}
\usepackage{amsmath} 
\usepackage{multirow} 
\usepackage{dirtytalk} 
\usepackage[utf8]{inputenc}
\usepackage[english]{babel}
\usepackage{fancyhdr}
\usepackage{lastpage}
\usepackage{comment}
\usepackage{caption}
\usepackage{subcaption}
\usepackage{adjustbox}
\usepackage{amssymb}
\usepackage{xcolor}
\usepackage{algorithm}
\usepackage{algpseudocode}

\algrenewcommand\algorithmicrequire{\textbf{Input:}} 
\algrenewcommand\algorithmicensure{\textbf{Output:}} 
\usepackage [english]{babel}
\usepackage [autostyle, english = american]{csquotes}
\MakeOuterQuote{"}

\usepackage{tikz}
\usepackage{graphics} 
\usepackage{graphicx}
\usepackage{epsfig} 
\usetikzlibrary{patterns}
\usetikzlibrary{trees}
\usetikzlibrary{shapes,arrows,shadows,positioning}
\usetikzlibrary{matrix, decorations.pathreplacing}
\pgfkeys{tikz/mymatrixenv/.style={decoration={brace},every left delimiter/.style={xshift=8pt},every right delimiter/.style={xshift=-8pt}}}
\pgfkeys{tikz/mymatrix/.style={matrix of math nodes,nodes in empty cells,left delimiter={[},right delimiter={]},inner sep=1pt,outer sep=1.5pt,column sep=2pt,row sep=2pt,nodes={minimum width=20pt,minimum height=10pt,anchor=center,inner sep=0pt,outer sep=0pt}}}
\pgfkeys{tikz/mymatrixbrace/.style={decorate,thick}}

\usepackage[acronym]{glossaries}
\makeglossaries
\newacronym[longplural={Information and Communication Technologies}]{ICT}{ICT}{Information and Communication Technology}

\newacronym{SC-PHY}{SC-PHY}{single carrier physical layer}
\newacronym{mmWave}{mmWave}{millimeter wave}
\newacronym{LoS}{LoS}{line-of-sight}
\newacronym{RL}{RL}{reinforcement learning}
\newacronym{AV}{AV}{autonomous vehicle}
\newacronym{OFDM}{OFDM}{orthogonal frequency division multiplexing}
\newacronym{CP}{CP}{cyclic prefix}
\newacronym{JCAS}{JCAS}{joint communication and  radar sensing}
\newacronym{ISAC}{ISAC}{integrated sensing and communication}
\newacronym{QAM}{QAM}{quadrature amplitude modulation}
\newacronym{MUSIC}{MUSIC}{multiple signal classification}
\newacronym{ESPRIT}{ESPRIT}{estimation of signal parameters via rotational invariance technique}
\newacronym{CS}{CS}{compressive sensing}
\newacronym{FPS}{FPS}{Fourier projection-slice}
\newacronym{LMMSE}{LMMSE}{linear minimum mean square error}
\newacronym{LS}{LS}{least square }
\newacronym{MMSE}{MMSE}{minimum mean square error}
\newacronym{CE}{CE}{channel estimation}
\newacronym{RMSE}{RMSE}{root mean square error}
\newacronym{TDD}{TDD}{Time Division Duplexing}
\newacronym{FDD}{FDD}{Frequency Division Duplexing}
\newacronym{UL}{UL}{Uplink}
\newacronym{DL}{DL}{Downlink}
\newacronym{BS}{BS}{Base Station}
\newacronym{MRC}{MRC}{Maximum Ratio Combining}
\newacronym{CSI}{CSI}{channel state information}
\newacronym{ACI}{ACI}{adjacent channel interference}
\newacronym[longplural={User Equipments}]{UE}{UE}{User Equipment}
\newacronym{3GPP}{3GPP}{3rd generation partnership project }
\newacronym{LTE}{LTE}{long term evolution}
\newacronym{AoA}{AoA}{Angle Of Arrival}
\newacronym{AoD}{AoD}{Angle Of Departure}
\newacronym{SFT}{SFT}{sparse Fourier transform}
\newacronym{ED}{ED}{Energy Detection}
\newacronym{ASK}{ASK}{Amplitude-shift keying}
\newacronym{PSK}{PSK}{Phase-shift keying}
\newacronym{ZCP}{ZCP}{Zadoff-Chu precoding}
\newacronym{ISM}{ISM}{industrial scientific and medical}
\newacronym{SNR}{SNR}{signal to noise ratio}
\newacronym{IDFT}{IDFT}{inverse discrete Fourier transform }
\newacronym{DFT}{DFT}{discrete Fourier transform }
\newacronym{IFFT}{IFFT}{inverse fast Fourier transform }
\newacronym{FFT}{FFT}{fast Fourier transform }
\newacronym{ISI}{ISI}{inter symbol interference}
\newacronym{ICI}{ICI}{inter carrier interference}
\newacronym{MIMO}{MIMO}{multiple input multiple output}
\newacronym{PAPR}{PAPR}{peak-to-average power ratio}
\newacronym{GLRT}{GLRT}{Generalized Likelihood Ratio Test}
\newacronym{i.i.d}{i.i.d}{independent and identically distributed}
\newacronym{SS}{SS}{Spectrum Sensing}
\newacronym{AWGN}{AWGN}{additive white Gaussian noise}
\newacronym{ZCT}{ZCT}{Zaddof-Chu transform}
\newacronym{HPA}{HPA}{high power amplifier}
\newacronym{RFI}{RFI}{Radio Frequency Interferences}
\newacronym{CPI}{CPI}{coherence processing interval}
\newacronym{MDP}{MDP}{Markov decision process}

\newacronym{MSE}{MSE}{mean square error}
\newacronym{BER}{BER}{bit error rate}

\newacronym{RADAR}{RADAR}{RAdio Detection And Ranging}
\newacronym{RadCom}{RadCom}{radar-communication}

\newacronym{RCS}{RCS}{radar cross section}
\newacronym{ITS}{ITS}{intelligent transportation systems}
\newacronym{JRC}{JRC}{joint radar-communication}
\newacronym{JCR}{JCR}{joint communications-radar}
\newacronym{C-ITS}{C-ITS}{cooperative intelligent transport systems}
\newacronym{DSRC}{DSRC}{dedicated short-range communication}
\newacronym{V2V}{V2V}{vehicle-to-vehicle}
\newacronym{V2I}{V2I}{vehicle-to-infrastructure}
\newacronym{V2X}{V2X}{vehicle-to-everything}
\newacronym{LRR}{LRR}{long range automotive radar}
\newacronym{JCR-AV}{JCR-AV}{joint communications-radar-autonomous vehicule}
\newacronym{DDDQN}{DDDQN}{double dueling deep Q-Learning network}
\newacronym{A2C}{A2C}{advantage actor-critic}
\newacronym{PPO}{PPO}{proximal policy optimization}
\newacronym{AoU}{AoU}{age of update}
\newacronym{RF}{RF}{reward function}
\newacronym{DQN}{DQN}{deep Q-learning}
\newacronym{SINR}{SINR}{signal-to-interference-plus-noise ratio}
\newacronym{BPSK}{BPSK}{binary phase-shift keying}
\newacronym{QSI}{QSI}{queue state information}
\newacronym{C-V2X}{C-V2X}{cellular vehicle-to-everything}
\newacronym{V2P}{V2P}{vehicle-to-pedestrian}
\newacronym{V2N}{V2N}{vehicle-to-network}
\newacronym{MORL}{MORL}{multi-objective reinforcement learning}  

\begin{document}

\bstctlcite{IEEEexample:BSTcontrol}

\title{\Huge Joint Adaptive OFDM and Reinforcement Learning Design for Autonomous Vehicles: Leveraging Age of Updates}

 \author{\authorblockN{Mamady Delamou\authorrefmark{1},
Ahmed Naeem \authorrefmark{2}, 
Hüseyin Arslan\authorrefmark{2}, 
El Mehdi Amhoud\authorrefmark{1}}\\
 \IEEEauthorblockA{\authorrefmark{1}College of Computing, University Mohammed VI Polytechnic , Benguerir, Morocco}\\
 \IEEEauthorblockA{\authorrefmark{2}Department of Electrical and Electronics Engineering, Istanbul Medipol University, Istanbul, 34810, Turkey}\\
 \{mamady.delamou, elmehdi.amhoud\}@um6p.ma, ahmed.naeem@std.medipol.edu.tr, huseyinarslan@medipol.edu.tr}

\maketitle
\begin{abstract}
Millimeter wave (mmWave)-based orthogonal frequency-division multiplexing (OFDM) stands out as a suitable alternative for high-resolution sensing and high-speed data transmission. To meet communication and sensing requirements, many works propose a static configuration where the wave's hyperparameters such as the number of symbols in a frame and the number of frames in a communication slot are already predefined. However, two facts oblige us to redefine the problem, (1) the environment is often dynamic and uncertain, and (2) mmWave is severely impacted by wireless environments. A striking example where this challenge is very prominent is autonomous vehicle (AV). Such a system leverages integrated sensing and communication (ISAC) using mmWave to manage data transmission and the dynamism of the environment. In this work, we consider an autonomous vehicle network where an AV utilizes its queue state information (QSI) and channel state information (CSI) in conjunction with reinforcement learning techniques to manage communication and sensing. This enables the AV to achieve two primary objectives: establishing a stable communication link with other AVs and accurately estimating the velocities of surrounding objects with high resolution. The communication performance is therefore evaluated based on the queue state, the effective data rate, and the discarded packets rate. In contrast, the effectiveness of the sensing is assessed using the velocity resolution. In addition, we exploit adaptive OFDM techniques for dynamic modulation, and we suggest a reward function that leverages the age of updates to handle the communication buffer and improve sensing. The system is validated using advantage actor-critic (A2C) and proximal policy optimization (PPO). Furthermore, we compare our solution with the existing design and demonstrate its superior performance by computer simulations. 
\end{abstract}

\IEEEoverridecommandlockouts
\IEEEpeerreviewmaketitle


\section{Introduction}
\PARstart{T}{he} upcoming wireless network generation is envisioned to enable ubiquitous and seamless communication, sensing, connectivity, and intelligence \cite{8869705}. Within this context, the emerging trend in 5G, 6G and beyond is to achieve higher spectral efficiency for unrestricted spectrum access to both sensing and communication systems. This concern has created significant interest in enabling coexistence between radar and communication systems in many applications such as autonomous vehicles \cite{6636787,7131242, covar_matrix_for_opt}. These vehicles are poised to revolutionize the traditional vehicular system by offering a diverse range of applications that enhance safety and efficiency within the transportation network. Within this context, two critical functions for autonomous cars come to the forefront: radar sensing and communication. The first vital function enables autonomous vehicles to identify and assess targets in their vicinity. By sensing, these vehicles can precisely detect objects, estimate their range, and determine their radial velocity. This capability plays a pivotal role in ensuring safe navigation, including collision avoidance, and precise object tracking, which functions are essential for reliable operations. On the other hand, communication permits the exchange of vital information in real-time. This cooperative data exchange allows for a deeper understanding of the surrounding environment, collaborative decision-making, and coordinated maneuvers, ultimately enhancing overall traffic safety and efficiency.\\
\indent The 5G automotive association is at the forefront of promoting the \gls{V2X} paradigm. This revolutionary approach allows vehicles to communicate not only with each other but also with various entities, expanding the connectivity possibilities beyond traditional \gls{V2V} communication \cite{Qualcomm}. The concept of \gls{V2X} encompasses a wide range of communication scenarios denoted by the `'X`', which could represent different entities, namely infrastructure, person, network, vehicle, etc.

\subsection{Litterature Review}
 In terms of vehicular communication, several wireless technologies have been proposed in the literature, namely, \gls{C-ITS} for intelligent transportation and \gls{DSRC} for \gls{V2X} communication. \gls{DSRC} is a wireless communication technology designed for direct, short-range communication between vehicles (\gls{V2V}) and between vehicles and infrastructure (\gls{V2I}), primarily for safety-critical and time-sensitive applications, such as collision avoidance, traffic management, and emergency vehicle communication \mbox{\textcolor{black}{\cite{10266768, 10556134}}}. It has been regarded as the conventional approach for vehicular communication, in line with IEEE 802.11p and operating on the 5.9 GHz \textcolor{black}{\cite{10560623}}.  An alternative to DSRC is known as \gls{C-V2X}, which supports V2V, \gls{V2I}, and network-based communication through cellular networks \textcolor{black}{\cite{10292930, 10049418}}. An evolution of C-V2X under 5G, named 5G NR V2X supports advanced V2X use cases with ultra-low latency, high reliability, and high throughput \textcolor{black}{\cite{10261272, 10533255}}. On the other hand, C-ITS encompasses a broader set of technologies and standards that facilitate cooperative behavior between road users (vehicles, pedestrians) and infrastructure \textcolor{black}{\cite{10220232, 10614832}}. It includes V2V, V2I, \gls{V2P}, and \gls{V2N} communications \textcolor{black}{\cite{10551825, 10437682}}. Although \gls{C-ITS} and \gls{DSRC} are efficient technologies, they are limited by their relatively low data rates of several megabits per second \textcolor{black}{\cite{electronics13142845}}, which do not meet the requirements for \gls{V2V} applications where data rates up to several gigabits per second may be \mbox{necessary \textcolor{black}{\cite{10606276}}}. To this end, utilizing \gls{mmWave} is the most convenient approach for high-bandwidth vehicle communication due to its wide bandwidth, which enables high-resolution radar sensing and fast data transmission. However, it still faces challenges as it is sensitive to environmental changes \cite{chu2022ai}.\\
\indent Several works have been proposed for vehicular communications \cite{9473593, 8463329, 9249538, 9952866}.
The work in \cite{9473593} reviews the key requirements, the state-of-the-art solutions, and the 5G radio roadmap; this is enhancing and offering new verticals to solve open challenges for vehicular networks.
In \cite{8463329}, the authors review the available wireless communication technologies to identify the most reliable option for enabling vehicle connectivity in Malaysia's transportation environment.
In \cite{9249538}, the authors investigate the temporal properties of vehicular visible light communication channels in dynamic traffic conditions. They focus on a two-dimensional temporal channel model considering time and signal propagation delay. The results show that the channels are time-variant nonstationary in high-density traffic with flat slow fading, while in low-density traffic, they can be stationary for short durations with uncorrelated reflectors and reduced multipath effects.
The authors in \cite{9952866} further analyze several different methods that have been proposed to overcome channel congestion issues in \gls{V2V} communication, which impede the timely transmission of periodic safety messages with increased traffic density.


In addition, in \cite{10594637, delamou2024interference}, they come up with a waveform design for \gls{ISAC} \gls{mmWave} vehicular networks to optimize communication performance while preserving sensing accuracy. Using \gls{OFDM}-based subcarrier-classified modulation method, they formulate the design problem into subcarrier division, sensing parameter computation, and \gls{BER} optimization. Another contribution comes from \cite{10485369}, where the authors present a survey of beam alignment techniques for mmWave \gls{V2X} communication, a critical enabler for autonomous vehicles, which demands high data throughput with low latency. They present some recent approaches, including beam sweeping, \gls{AoA}/\gls{AoD} estimation, and black-box optimization, and discuss future research challenges toward enhancing reliability and efficiency in mmWave vehicular networks.
In \cite{kumari2015investigating, kumari2017ieee}, the authors investigate a unified framework for long-range automotive radar based on the IEEE 802.11ad frames. A low complexity \gls{JRC} receiver for bi-static automotive radar in conjunction with \gls{V2V} communications is presented in \cite{dokhanchi2019mmwave}. 
Furthermore, an opportunistic radar system that exploits probing signals generated during sector-level scanning for range and velocity estimation is proposed in  \cite{grossi2018opportunistic}. However, this system encounters challenges in achieving accurate radial velocity due to the brief duration of the probing signals.\\
\indent A common drawback in the above studies is that the waveform parameters such as the count of frames within a communication slot, the modulation scheme for communication, and the number of symbols in a frame are static. Therefore, the work introduced in \cite{kumari2019adaptive} suggests dynamically choosing the number of frames in a \gls{CPI} that can compensate for communication and sensing resolution for vehicular applications. However, the system needs complete information about the environment. Moreover, most of the previously presented work utilize the classical Friis communication model, which is based on attenuation. In this model, attenuation refers to the loss of signal strength as it travels from the transmitter to the receiver. For a more pragmatic approach, we adopt \gls{V2V} channel model where the communication link includes a factor called the channel blocker coefficient, which represents the level of obstruction in the \gls{LoS} link.\\ 
\indent In addition, none of the prior studies take into account the dynamic nature and uncertainties associated with the information and environment. To fill this gap, the authors in \cite{chu2022ai} proposed a framework that optimizes the waveform structure in response to the unpredictable characteristics of the surrounding environment, while the \gls{AV} is in motion. They initially formulate the concern based on \gls{MDP} and subsequently, validate the system by employing the Q-learning algorithm and its various extensions. To our knowledge, this was the first approach that took into account the dynamic of the environment during the learning process.
\subsection{Contributions}
In this work, we suggest an optimization approach by concurrently introducing \gls{QSI} along with \gls{CSI}. \gls{CSI} conveys information regarding transmission, while \gls{QSI} indicates the urgency of the traffic. In addition, compared to \cite{chu2022ai}, instead of focusing only on the \gls{QSI} to evaluate the communication performance, we introduce the \gls{AoU}. It refers to the amount of time a packet has spent in the buffer since its arrival, which gives a deep insight into the traffic. Additionally, we explore an adaptive \gls{OFDM} system that adjusts modulation schemes based on the overall state of the system. The proposed scheme is compared with the pioneering works, showing enhanced and better performance for communication and sensing.\\
\indent The primary contributions of this work are outlined as follows:
\begin{itemize}
   \item \gls{AoU} has been introduced in the literature for energy efficiency as in \textcolor{black}{\cite{9794610, 10501858}} and for updated information transmissions as mentioned in \textcolor{black}{\cite{10681887,8937801,9628120}}. In our work, we leverage \gls{AoU} to measure the freshness of information in \gls{V2V} link. We achieve a better trade-off between communication latency and velocity estimation by incorporating \gls{AoU} into the \gls{RF}. 
   
   
   \item The problem is formulated as a \gls{MORL} because the objective function to optimize is a vector containing the queue length, the radar accuracy, and the dropped packets rate. Since policy gradient methods for instance \gls{PPO} and \gls{A2C} require a scalar objective function to compute gradients effectively, we linearly scalarized the objective function to obtain a single objective. This operation permits the agent to learn the behavior of a \gls{V2V} link based on weighted \gls{RF}, where the obstruction of the \gls{LoS} link occurs with a certain probability. The weighted \gls{RF} allows a trade-off for stable communication and accurate sensing.
   \item We introduce an adaptive action-state framework that combines modulation scheme selection in conjunction with \gls{OFDM} frame size in the action set while incorporating the \gls{SINR} into the state representation. This dual enhancement enables dynamic adjustments to data transmission parameters based on channel conditions, significantly reducing packet loss in adverse scenarios and improving sensing and communication reliability.

   We exploit the adaptive \gls{OFDM} technique to adjust the modulation schemes based on real-time channel conditions learned by the \gls{RL} agent. 
   
   \item Finally, through simulations, we compared our scheme to the introductory work presented in \cite{chu2022ai}, where the adaptive modulation and \gls{AoU} are not under consideration. As shown in the results, our approach not only enhances data reliability but also improves sensing accuracy in any different channel conditions.   
\end{itemize}

To the best of our knowledge, we are not aware of any work that addresses the \gls{ISAC} requirements problem in a dynamic \gls{V2V} network using metrics including information freshness, average dropped packets, average effective data rate, and velocity accuracy.\\
\indent The remainder of the work is organized as follows: In Section II, we introduce the proposed system model followed by the problem statement in Section III. Afterwards, we describe the optimization process using the \gls{RL} agent in section IV. Simulation results and discussions are presented in Section V. Finally, Section VI concludes the work and sets forth our perspectives.
\section{System Model}

\subsection{Communication Channel Modeling}
\subsubsection{Attenuation Modeling}

 \begin{figure}[t!] 
    \centering
    \includegraphics[width=3.5in]{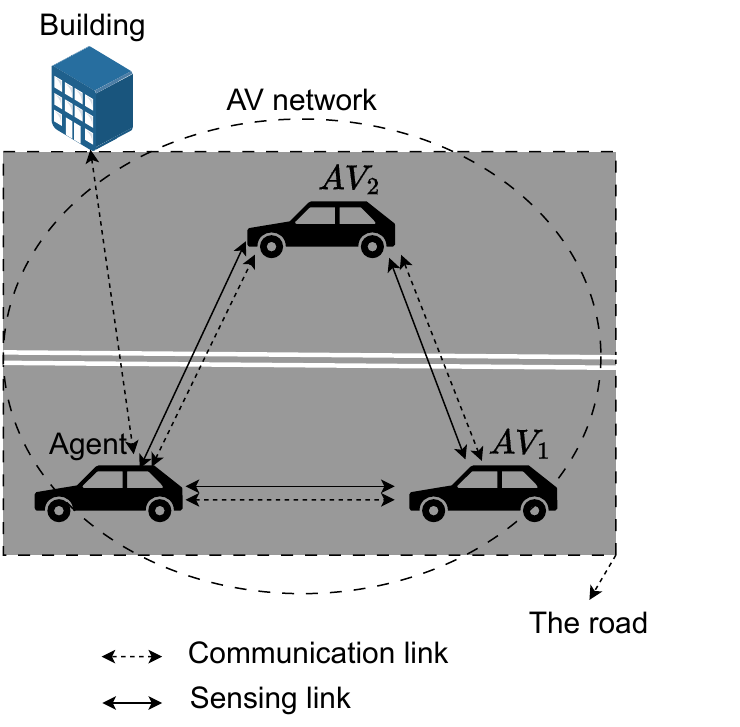}
    \caption{AV network}
    \label{system_mod_1}
\end{figure}
We consider a set of \glspl{AV} moving in the same direction on the road and equipped with \gls{ISAC} capabilities, meaning that they can establish a communication link with other \gls{AV}s and sense the surrounding objects. By considering that, each \glspl{AV} has a transmitter $T_X$ and receiver $R_X$, and it can establish communication through V2V links on the mmWave frequency band, exploiting \gls{OFDM} mode of IEEE 802.11ad. At each communication round, an \gls{AV} has communication links with the other \glspl{AV} to send and receive information, to report an emergency braking, for instance, as mentioned in \textcolor{black}{\cite{9594815}}. At the same time, for sensing purposes, it processes the reflected signals coming from the AVs and other objects such as buildings and trees along the road, to estimate their range and radial velocity as depicted in Fig. \ref{system_mod_1}. We denote by $\mathcal{T}$ = $\{1,2,\dots,T\}$, \mbox{$\mathcal{R}$ = $\{1,2,\dots,R\}$}, \mbox{$\lvert \mathcal{T} \rvert$ = $T$}, \mbox{$\lvert \mathcal{R} \rvert$ = $R$}, the sets containing the number of  $T_X$, $R_X$ and their cardinalities, respectively. We consider a total bandwidth $B_{\text{w}}$ which can be divided into $N$ bands with central frequencies $f_0$,$f_1$,\dots$,f_{N-1}$ such that $\Delta f=\frac{B_{\text{w}}}{N}$, where $N$ denotes the number of subcarriers. An \gls{OFDM} symbol is a packet of $N$ modulated data transmitted at the same time on $f_0$,$f_1$\dots$f_{N-1}$.\\ 
\indent As the \gls{CSI} describes how the signal behaves as it travels from the transmitter to the receiver, including the effects of fading, interference, and noise, by tracking it, the system can switch between various modulation schemes, spanning from \gls{BPSK} with one bit per symbol to more complex schemes like 64-\gls{QAM} with six bits per symbol. This adaptation enables the system to optimize data transmission rates and reliability in response to changing channel characteristics. In favorable channel conditions, higher-order modulations such as 64-QAM can be utilized to achieve higher effective data rates. However, in harsher environments, adaptive OFDM switches to more robust modulations, such as \gls{BPSK}, to reduce the packet error rate.\\ 
\indent Due to the continuous movement of vehicles, the wireless communication channel undergoes rapid and frequent changes. Therefore, maintaining a stable and reliable connection is challenging. In particular, a \gls{mmWave} communication, while offering high effective data rates and capacity, is highly sensitive to obstructions and changes in the surrounding environment. One of the critical challenges involves the necessity for beam alignment. Precisely, \glspl{mmWave} have narrow beamwidths, and a precise alignment between the transmitter and receiver antennas is essential for optimal signal strength and data transmission. However, in a dynamic vehicular environment, achieving and maintaining this alignment can be challenging, as vehicles are constantly changing their positions and orientations. These significant topological variations necessitate the adoption of time-slotted communications. Thoroughly, we denote by $s$ the time slot dedicated to data transmission, and by $\Tilde{s}$ the scheduling time, which is dedicated to beam alignment. When a scheduling process is triggered at the beginning of a given time slot, the slot is divided into alignment time and effective data transmission time, followed by the other data transmission slots until new scheduling is requested.\\
\indent To accurately model the propagation characteristics, a \gls{V2V} link is set up at carrier frequency $f_c$ = \mbox{60 GHz} mmWave frequency along with the effects of blockages, such as obstacles and physical obstructions. The log-distance path-loss model provides a suitable foundation, accounting for signal attenuation over distance. We denote by $T_X^i$ and $R_X^j$, the $i$th transmitter and the $j$th receiver, respectively, with $i \in \mathcal{T}' 	\subseteq \mathcal{T}$ and $j \in \mathcal{R}' 	\subseteq \mathcal{R}$. For given $T_X^i$, $R_X^j$, $\mathcal{R}'$ is the set of the receivers linked to $T_X^i$ and $\mathcal{T}'$ is the set of transmitters linked to $R_X^j$. Based on the model proposed in \cite{8187182, 4357009}, the channel gain $g^c_{ij}$ between $T_X^i$ and $R_X^j$ is given by
\begin{equation}\label{channel_gain}
g_{ij}^c=10 \delta_{ij} \log _{10}\left(s_{ij}\right)+\beta_{ij} + 15 s_{ij} /1000,
\end{equation}
where $\delta_{ij}$, $s_{ij}$, $\beta_{ij}$ and $15 s_{ij} /1000$  represent the path-loss exponent in land mobile communications \cite{4357009}, the distance between $T_X^i$ and $R_X^j$ in meters, a constant, and the atmospheric attenuation of 15~dB/km at 60 GHz, respectively. It is worth noting that $\delta_{ij}$ and $\beta_{ij}$ are obtained by better fitting. The model suggested in \cite{4357009} was generalized by the works proposed in \cite{8187182}, where the values of $\delta_{ij}$ and $\beta_{ij}$ were estimated for a general case when the number of obstructing vehicles is more than three.\\
\indent The \gls{CSI} of the entire system is estimated by computing the average \gls{CSI} of each link within the system. Due to the mobility of vehicles, the channel gain between any two vehicles undergoes dynamic changes over time slots, resulting in a time-dependent channel gain denoted as $g_{ij}^c(s)$. Consequently, at each time slot $s$, the global state of the channel, as perceived from $R_X^j$ perspective, is given by
\begin{equation}
H_j(s) = \frac{H_{ij}(s)}{\lvert \mathcal{R}' \rvert},
\end{equation}
where $H_{ij}(s) = g_{ij}^c(s)$ if link $l_{ij}$ exists.\\

\subsubsection{Antennas Alignment and Transmission Rate}

We consider a two-dimensional sectored antenna, taking into account the boresight direction and the half-power beamwidth. The probability of misalignment affecting a desired link $l_{ij}$ can vary based on multiple factors, including the vehicles' relative speed in $l_{ij}$, the main-lobe width of $T_X^i$ and $R_X^j$, and the duration of the scheduling process. Additionally, the chosen beamwidths determine whether signals from links other than $l_{ij}$ affect the mainlobe of the receiver antennas, significantly impacting the measured \gls{SINR}. The cost associated with the alignment $\tau_{ij}(s)$ can be quantified as in \cite{7959165}: 
\begin{equation}
\tau_{ij}(s) \triangleq \tau_{ij}\left(\varphi_{i}(s), \varphi_{j}(s)\right)=\frac{\psi_i \psi_j}{\varphi_{i}(s) \varphi_{j}(s)} T_p,
\end{equation}
where $\psi_{i}(s)$ and $\psi_{j}(s)$, $\varphi_{i}(s)$ and $\varphi_{j}(s)$ denote the sector-level beamwidths and the half-power beamwidths of the link $l_{ij}$, between $T_X^j$ and $R_X^j$, respectively. $T_p$ represents the pilot transmission interval. Given that $\tau_{ij}(s)$ should be less than $s$, in addition to constraints related to operational array antenna and sector-level beamwidths, the beamwidth of $T_X^i$ and $R_X^j$, $\varphi_{i}(s)$ and $\varphi_{j}(s)$ must verify that: 
\begin{equation}
\varphi_{i}(s)\varphi_{j}(s) \ge \frac{T_p}{s}\psi_{i}(s)\psi_{j}(s).
\end{equation}
As a result, the maximum achievable effective data rate at the end of $s$, denoted as $\mu_{ij}(s)$, for $l_{ij}$, is contingent on whether beam alignment occurs during the specific time slot and to the SINR measured at $R_X^j$. The effective data rate for a time slot $s$, which aligns with the duration of $s$ over which beam alignment is performed can be expressed as follows \cite{7959165}:
\begin{equation}\label{effective_data_rate}
\mu_{ij}(s)=\left(1-\frac{\tau_{ij}(s)}{s}\right) B_{\text{w}} \log _2\left(1+\operatorname{\eta}_j(s)\right),
\end{equation}
where the $\eta_j$ is the \gls{SINR} at $R_X^j$, given by:
\begin{equation}\label{sinr}
\operatorname \eta_j(s) =  \frac{\rho_i g_{i}^{a}(s) g_{ij}^c(s) g_{j}^{a}(s)}{\sum_{\substack{k \in \mathcal{T'} \subseteq \mathcal{T} \\   k \neq i}} p_k g_{k}^{a}(s) g_{kj}^c(s) g_{j}^{a}(s) + N_0 B_{\text{w}}},
\end{equation}
with $\rho_i$ the transmission power of $T_X^i$,  $g_{i}^{a}(s)$ and $g_{j}^{a}(s)$ denoting the antenna gains at $T_X^i$ and $R_X^j$, respectively. The term \mbox{$I = p_k g_{k}^{a}(s) g_{kj}^c(s) g_{j}^{a}(s)$} represents the interference at the $R_X^j$ from all the $T_X^k$, $k \in \mathcal{T'}$, $k \neq i$, whereas $N_0$ is the Gaussian noise power density (dBm/Hz).\\

\subsubsection{Queue and Delay Modeling}
After a communication is initiated, the transmitter starts sending data packets, and the receiver checks the integrity and correctness of all the packets and sends an acknowledgment back to the sender. At the end, if the sender receives the acknowledgment within a specified timeout period, it considers the packets as successfully delivered. If the acknowledgment is not received within the timeout period, the sender assumes that the packets were lost or corrupted and takes appropriate actions, for instance, their retransmission.\\
\indent We denote by $\lvert \mathcal{P} \rvert$ the size of a data packet and we assume that $q_j(s)$ is the queue length in number of packets at $R_X^j$ at the end of time slot $s$.  We denote by \mbox{$\mathcal{\lambda_{T'}}(s)$ = $[\lambda_1(s), ..., \lambda_{T'}(s)]$} the vector representing the random packet arrivals at $R_X^j$ from all $T_X^i$ linked to $R_X^j$ at the end of time slot $s$. We also hypothesize that every entry $\lambda_i(s)$ in $\mathcal{\lambda_{T'}}(s)$, $\forall i \in \{1, . . ., T'\}$, is independently and identically distributed (i.i.d.) across time slots and follows a Poisson distribution, with average \mbox{$\lambda = \mathbf{E}[\lambda_i(s)] = \lambda_j$}. The queue dynamics for $R_X^j$ are given as in \cite{7959165} by:
\begin{equation}\label{queue_update}
q_j(s)=\min \left\{\left(q_j(s-1)-\frac{\mu_{*,j}(s) s}{\lvert \mathcal{P} \rvert}\right)^{+} + \lambda_j(s), q_{\max }\right\},
\end{equation}
with $x^+ \triangleq \max\{x, 0\}$
Hence, we calculated the number of dropped packets as follows: At the end of each time slot $s$, the total number of packets available at $R_X^j$ is equal to the count of packets in the buffer at the beginning of $s$, and the count of incoming packets $\lambda$, from which we subtract the number of packets transmitted by $R_X^j$ based on the channel capacity. There is a drop if the resulting number of packets is greater than $q_{max}$. From this, we derived the number of dropped packets $d_j$ at the end of $s$ as follows:

\begin{equation} \label{drop_paquets}
d_j(s)=\left\{\left(q_j(s-1)-\frac{\mu_{*, j}(s) s}{\lvert \mathcal{P} \rvert}\right)^{+} + \lambda_j(s) - q_{\max }\right\}^+,
\end{equation}
with $q_{max}$ and $\mu_{*,j}$ denote the maximum buffer size and the total effective data rate available at $R_X^j$, respectively.\\
\indent Upon its arrival to the queue, a packet can be delivered or dropped. In addition, a packet is dropped if, upon its arrival, the queue is already full. Every dropped packet is no longer recoverable, so we define a packet as successfully delivered if it is not dropped upon reaching its destination. 

\subsection{Sensing Channel Modeling}

Given the \gls{ISAC} hypothesis, the agent should be continuously sensing the surrounding environment to avoid collisions. Consequently, we assume that there is sensing whenever a communication process is triggered. In addition, jamming data is sent in the intervals where there is no data transmission. Furthermore, a communication time slot is composed of $\bar{N}$ \gls{OFDM} frames.
\begin{figure}[t!] 
\centering
\includegraphics[width=3.5in]{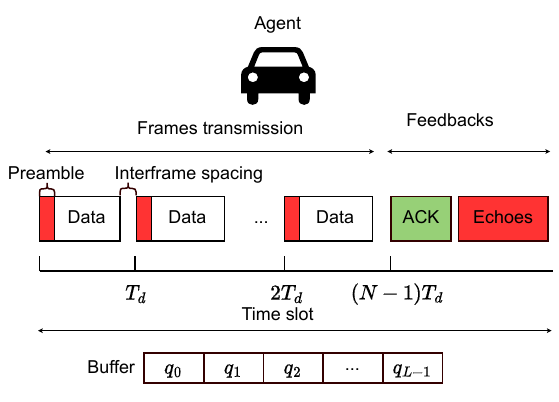}
\caption{Structure of the communication where frames
are arranged at multiples of $T_d$.}
\label{data_structure}
\end{figure}
Since the AV is moving, the environment around it changes considerably, in addition to the fact that \glspl{mmWave} are very sensitive. We model this instability by a packet error rate distribution $\mathbf{B'}$ and the link obstruction probability vector $\mathbf{P'}$. When a frame is lost, all the packets within the frame are also lost.\\
\indent The IEEE 802.11ad \gls{SC-PHY} frame is composed of two blocks, a fixed size preamble, which contains multiple Golay sequences, and a data field of varying length that carries the payload as shown in \mbox{Fig. \ref{data_structure}}. In addition, the good ambiguity function property of Golay sequences exhibits an ideal auto-correlation, making it ideal for use in radar functions, e.g., range estimation and multitarget detection \cite{kumari2017ieee}. Nonetheless, it exhibits high sensitivity to Doppler shifts, which results in low resolution of velocity. The higher the number of frames, the better the velocity resolution. Therefore, multi-frame has been introduced to improve the velocity estimate \cite{kumari2019adaptive}.\\
\indent By denoting $v_{max}$, the maximum target’s radial velocity that can be detected, the maximum Doppler shift $\Delta f_{max}$ caused by the objects is such that $\Delta f_{max}=\frac{2v_{max}}{\omega}$, with $\omega$ representing the wavelength \cite{delamou2022efficient}.\\
We consider $x[n]$ as the \gls{OFDM} symbols sequence of the $n$th frame with $K_s$ symbols. The discrete-time complex baseband representation of the transmitted signal is expressed as \cite{delamou2022efficient, Braun2014OFDMRA}:
  \begin{equation}\label{frame_resp}
      x[n]=\sum^{K_s-1}_{l=0} \sum^{N-1}_{k=0} o_{k,l} 
      e^\frac{{j2\pi k(n-lN_{s})}}{N},
 \end{equation}
 with $o_{k,l}$, the outputs of the \gls{IFFT} block in the \gls{OFDM} transmitter.\\
According to \cite{Braun2014OFDMRA, delamou2022efficient}, a large sub-carrier distance increases the de-orthogonalizing effect of the \gls{OFDM} subcarriers. Therefore, to keep the orthogonality, the velocity should be such that 
\begin{equation}\label{vmax}
    v_{max} \ll \frac{c\Delta f }{2f_c},
\end{equation}
which depends on the wave parameterization. By assuming that, the discrete-time complex representation of the received signal is expressed as: 
\begin{equation}
\begin{gathered}
y[n,l]=\sum^{N_t-1}_{c=0} \sum^{K_s-1}_{l=0} \sum^{N-1}_{k=0}  g_c o_{k,l}
 e^{j2\pi \frac{kn}{N}}
 e^{-j2\pi k \Delta f \tau_c}
 e^{\frac{j 2 \pi f_{D_{c}} lN_s}{N \Delta f}}
 \\
 \times e^{j\varphi_c}
       + z[n,k],
\end{gathered}
\end{equation}

where: 
 \begin{itemize}
     \item $g_c$ is the communication complex channel gain as described in (\ref{channel_gain}) and $z_n[n,k]$ is the complex white Gaussian noise with zero mean and variance $\sigma_c^2$, denoted by $\mathcal{N}_C\left(0, \sigma_c^2\right)$.
     \item$ \tau_c$ is the back and forth delay from the agent to the target and from the target back to the agent, and $\tau_c=\frac{2\bar{d}}{c_0}$. Where $\bar{d}$ is the distance between the agent and the target and $c_0$ is the speed of light.
     \item $f_{D_c}$ is the Doppler shift due to the speed of the target, and $f_D = \frac{2v}{c_0}f_c$. 
     \item $N_s=N+N_{cp}$, where $N_{cp}$ is the number of complex symbols transmitted as cyclic prefixes. 
     \item $\varphi$ is a random rotation phase introduced when the signal hits the target.

 \end{itemize}

As described in \cite{10011222, delamou2022efficient}, a distance $\Delta d$ (respectively velocity $\Delta v$) is called the radar resolution if it is the lowest distance (respectively velocity) such that two targets positioned at $d$ and $d+\Delta d$ (respectively moving at velocities $v$ and $v+\Delta v$) can still be distinguished, i.e., 
\begin{equation}\label{radar_resolution}
      \Delta d=\frac{c}{2 N \Delta f}, ~\mathrm{and}~\Delta v=\frac{c}{2 \bar{N} f_c T_s},
 \end{equation}
 where $\bar{N}$ represents the number of frames employed for the velocity estimation process. It is important to recall that achieving good distance resolution is relatively easy in most applications \cite{chu2022ai}. However, improving velocity resolution is much more challenging due to the sensitivity to the Doppler effect and constraints on augmenting the number of frames. In addition, a large number of frames can congest the communication link. Therefore, in this work, we focus only on the velocity resolution estimation. The precision of velocity measurement is reflected in the root mean square error, which is contingent on the \gls{SINR} of the received radar signal, and can be expressed as follows \cite{chu2022ai}:
\begin{equation}\label{velo_reso}
\varrho=\frac{c}{2 \bar{N} T_d f_c \sqrt{2\eta}}.
\end{equation}

\section{Problem Formulation}

\begin{figure}[t!] 
\centering
\includegraphics[width=3.5in]{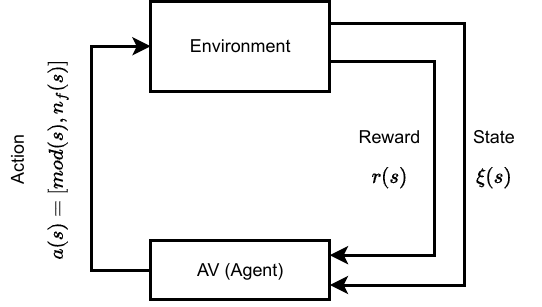}
\caption{Reinforcement learning scheme}
\label{rl_scheme}
\end{figure}

Let us consider two vehicles $\text{V}_i$ and $\text{V}_j$ in the network, connected by the link $l_{ij}$, where $\text{V}_i$ serves as the transmitter. From a sensing perspective, $\text{V}_j$ is a target to be detected, while from a communication perspective, it is the receiver. 
The link $l_{ij}$ can undergo fading depending on whether there are obstacles (for example other vehicles $\text{V}_k$ with $k \ne \{i,j\}$) blocking the signal or not.
We consider four levels of fading, called blocking levels or blocking status and we denote them $\mathbf{B} = (b_i)_{i \in \{0, \dots, 3\}}$.
$b_0$ represents the \gls{LoS} scenario, where there are not any obstacles between $\text{V}_i$ and $\text{V}_j$. $b_1$, $b_2$, and $b_3$ represent scenarios denoted respectively by "1V", "2V", and "3V", where one, two, and three vehicles different from $\text{V}_i$ and $\text{V}_j$, block the signal, respectively.
In addition, the blocking levels implicitly impact the channel gain defined in Eq. (\ref{channel_gain}), since the values of $\delta_{ij}$ and $\beta_{ij}$ dependent on $b_i$ \cite{8187182, 4357009}.
We note $\mathbf{P} = (p_i)_{i \in \{0, \dots, 3\}}$, the probabilities associated with each blocking level, respectively. The blocking level $b \in \mathbf{B}$ is selected based on the probability vector $\mathbf{P}$, noted as $b \sim \mathbf{P}(\mathbf{B})$. 
For instance, for $B = [0.4, 0.3, 0.2, 0.1]$, 40\% of the communications are \gls{LoS}, 30\% are "1V", 20\% are "2V", and 10\% are "3V". Therefore, based on the ratios of "LoS", "1V", "2V", and "3V" communications, the channel can be \textit{strong}, \textit{normal} or \textit{poor}, describing the reliability of the transmission.\\
\indent On the other hand, the link $l_{ij}$ is also characterized by the packet error rate $\mathbf{B'} = (b'_i)_{i \in \{0, \dots, 2\}}$, where $b'_0$, $b'_1$, and $b'_2$ indicate the current packet error rates for \textit{strong}, \textit{normal}, and \textit{poor} links, respectively.
As in conventional communication the packet error rate is required to be less than 1\% \cite{6560009,chu2022ai}, we pose that $b'_0 \ll 1\%$ and $b'_1 = 1\%$ and $b'_2 \gg 1\%$. Similarly as blocking levels, we denote $\mathbf{P'} = (p'_i)_{i \in \{0, \dots, 2\}}$, the probability vector associated to each packet error rate. The packet error rate $b' \in \mathbf{B'}$ is selected based on the probability vector $\mathbf{P'}$, noted as $b' \sim \mathbf{P'}(\mathbf{B'})$.\\
\indent As the system must handle communication and sensing tasks, our primary objective is to optimize performance by carefully balancing data transmission efficiency and sensing accuracy. To achieve this, we need a comprehensive \gls{RL} agent that considers both aspects. Such a system can be viewed as a finite \gls{MDP} described by the 4-tuple $\mathcal{M} = (\mathcal{S}, \mathcal{A}, \mathcal{Z}, \mathscr{P})$, where the state set $\mathcal{S}$ and action set $\mathcal{A}$ are both composed of a finite number of states $\xi$ and actions $a$, and $\mathcal{Z}$ represents the set of rewards $r$. Moreover, $\mathcal{M}$ involves transition-reward distributions that are defined at a time slot $s$ as \mbox{$\mathscr{P}(\xi(s+1) = \xi', ~r(s+1) = r' ~|~ \xi(s) = \xi_s, ~a(s) = a_s)$}. The system state and action sets can be described as follows:
\begin{itemize}
    \item State Set: It is all possible states that the system can be within the environment and comprises two variables, the \gls{QSI}, represented by the number of packets currently in the buffer $q$, and the \gls{SINR} $\eta$. It is a tuple expressed as:
    \begin{equation}\label{state_space}
        \mathcal{S}=\{(q, \eta): q \in\{0, \ldots, q_{max}\}\}.
        \end{equation}
In this work, we focus on managing the quality of links in the vehicular network rather than the vehicles' resources. Eq. (\ref{state_space}) represents the system state $\mathcal{S}$, which is used to describe the quality of the link between two vehicles. This quality is crucial for triggering high data rate transmission and efficient sensing. The system state $\mathcal{S}$ incorporates \gls{SINR} $\eta$ to assess the reliability of the link. A high \gls{SINR} indicates a low packet error rate, which supports efficient communication. However, the \gls{SINR} alone is not sufficient. Therefore, the system also considers \gls{QSI} $q$ to ensure low packet drop rates due to queue overflow. By combining \gls{SINR} and \gls{QSI}, the state vector $\mathcal{S}$ provides a comprehensive view of the link quality, balancing reliability and queue management. 
        
    \item Action Set: It includes all possible actions the agent can take when in any given state and comprises two variables, the modulation size $\bar{m}$, i.e., the number of bits in a communication symbol, and the number of frames $\bar{n}$. The set of possible actions can be described as follows:
    \begin{equation}
        \mathcal{A}=\{(\bar{m}, \bar{n}):  \bar{m} \in\{1, 2, 4, 6\} ; \bar{n} \in\{1, \ldots, \bar{N}\}\},
    \end{equation}       
\end{itemize}

\indent Let us consider a time slot $s$, the random variables for action, state, and reward are represented as $a(s) \in \mathcal{A}$, $\xi(s) \in \mathcal{S}$, and $r(s) = r\left(\xi(s), a(s)\right)$, respectively. The cumulative sum of all rewards acquired after time step $s$ with a discount factor $\gamma \in (0,1]$ for a sequence of length $L$ is given by:
\begin{equation}\label{cum_reward}
    C(s) = \sum_{k=0}^{L-s} \gamma^k r(s+k+1).
\end{equation}
The action-value function under a specific policy $\pi$ is $\mathbb{E}_{\pi} [C(s) \mathrm{~}|\mathrm{~}\xi(s) = \xi_s, a(s) = a_s]$. The primary objective is to learn a policy $\pi^*$ that maximizes the expected return:
\begin{equation}\label{optimal_policy}
    \pi^* = \arg\max_{\pi}\mathbb{E}_{\pi}[C(s)].
\end{equation}
Given the dynamic and uncertain nature of the environment where factors such as packet drop probability and data arrival rate impact system performance, the agent has no prior knowledge of these variables. To address this challenge, \gls{RL} offers a promising approach by enabling the system to acquire an optimal policy by interacting with its environment as depicted in Fig. \ref{rl_scheme}.\\ 
\indent We suppose that at the end of the time slot $s$, the system moves to state $\xi(s)$ and takes action $a(s)$, and we note $q(s)$, $\varrho(s)$, $\nabla(s)$ and $d(s)$, the count of packets in the queue, the radar accuracy, the successfully delivered effective data rate, and the count of discarded packets, respectively. Let us assume that the action $a = (\bar{m},\bar{n})$ is returned as the optimal action at a given time $s$, where $\bar{m}$ is the modulation scheme and $\bar{n}$ the number of frames. Consequently, as an \gls{OFDM} symbol comprises \( N \) symbols, each subcarrier transmits \(\bar{m}\) bits ($\frac{\bar{m}}{8}$ bytes). The total bytes in an \gls{OFDM} symbol is $\frac{N \times \bar{m}}{8}$. Considering each \gls{OFDM} symbol as a frame, and knowing \(\bar{n}\) as the number of frames during $s$, the total number of bytes transmitted across all frames during \( s \) is given by $\frac{N \times \bar{m} \times \bar{n}}{8}$. Therefore, for two vehicles $\text{V}_i$ and $\text{V}_j$, the successful transmit data rate \(\nabla\) in terms of the number of packets, can be expressed as:
\begin{equation}\label{data_rate}
    \nabla (s) = (1-b') \frac{N\times\bar{m}\times\bar{n}}{8\lvert \mathcal{P} \rvert} \le \mu_{ij},
\end{equation}
where $b'$ is the packet error rate, $\lvert \mathcal{P} \rvert$ the size of one data packet in bytes, and $\mu_{ij}$ is the maximum data rate between  $\text{V}_i$ and $\text{V}_j$ during the time slot $s$ as described in Eq. (\ref{effective_data_rate}).\\
\indent One design of the \gls{RF} as formulated in \cite{chu2022ai} can be expressed as:
 \begin{equation}\label{reward_fun1}
r(s) = r\left(\xi(s), a(s)\right)=-\left(\omega_1 q(s) + \omega_2 \varrho(s) + \omega_3 d(s)\right),
\end{equation}
where $\omega_1$, $\omega_2$, and $\omega_3$ are the weights.\\
\indent The optimization problem consists of selecting, for each communication round, the number of frames $\bar{n}$ in the time slot and the modulation scheme $\bar{m}$ that would increase velocity resolution and improve communication through better management of the queue, high data rate and low dropped packets. In typical \gls{RL} jargon, the agent’s goal is to find a policy $\pi$ (a mapping from states to actions) that maximizes the cumulative reward over time by taking actions that move it into favorable states.

\section{\gls{PPO} and \gls{A2C} for \gls{V2V} Link Optimization}

\indent \gls{RL} has emerged as a highly effective approach for managing and adapting to dynamic and changing environments. In the \gls{V2V} environments, changes in channel states, caused by factors like obstruction due to cars, and buildings, must be continuously monitored and learned. \gls{RL} enables an agent to learn and adapt through interactions with the environment, without requiring prior complete knowledge. The \gls{RL} agent learns to manage these changes by analyzing the data queue in the buffer and resolving velocity issues. This process involves continuously updating its policy to optimize performance. For instance, the agent learns to choose the appropriate number of frames to transmit in each communication round to ensure efficient communication and accurate sensing. To assess the effectiveness of our proposed system model, we validate it using two advanced \gls{RL} algorithms: \gls{PPO} and \gls{A2C}.\\
 \indent This work is an approach to demonstrate how, by introducing adaptive OFDM and by leveraging the \gls{AoU} we can improve the reward defined in \mbox{Eq. (\ref{reward_fun1})}. The performance evaluation criteria for communication are the number of packets in the buffer, the number of discarded packets (lost packets), and the number of successfully delivered packets. In contrast, the sensing is evaluated through the velocity resolution expressed in \mbox{Eq. (\ref{velo_reso})}.


\begin{algorithm}[!t]
   \caption{V2V Link Optimization}\label{pseudo_algo}
    \begin{algorithmic}[1]
    \Require  $q = 0$, blocker list $\mathbf{B} = [\text{"LoS", "1V","2V", "3V"}]$, $\mathbf{B'} = [10\%, 1\%, 0.3\%]$
    \Ensure Average values of the weighted reward $r$, queue length $q$, velocity resolution $\varrho$, data rate $\nabla$, and dropped packets $d$. 

    \Function{Env}{$\xi$}
        \State $\bar{m}, \bar{n} \gets \text{model.predict}(\xi)$
        \State Knowing the action $a = (\bar{m}, \bar{n})$, compute:
        \State $q$ as in Eq. (\ref{queue_update})
        \State $d$ as in Eq. (\ref{drop_paquets})
        \State $\varrho$ as in Eq. (\ref{velo_reso})
        \State $b' \sim \mathbf{P'}(\mathbf{B'})$
        \State $\nabla$ as in Eq. (\ref{data_rate})
        \State $r$ as in Eq. (\ref{reward_fun2})
        \State $b \sim \mathbf{P}(\mathbf{B})$
        \State Compute $g^c_{ij}$ as in Eq. (\ref{channel_gain}) and $\eta$ as in Eq. (\ref{sinr})
        \State \Return $r, q, \varrho, \nabla, d, b$
    \EndFunction
    \State Randomly select $(b, b') \in (\mathbf{B}, \mathbf{B'})$
    \State Compute $g^c_{ij}$ as in Eq. (\ref{channel_gain}) and $\eta$ as in Eq. (\ref{sinr})
    \State $\xi \gets (q, \eta)$
    
 \ForAll{$e \in \text{EPISODE}$}
    \Statex \hspace{0.5em} \textbf{Training Step:}
    \State $\text{model} \gets \text{A2C}(\text{Env}(\xi))$ \Comment{Same for PPO}
    \State model.learn(ITERATION) \Comment{Train on \textbf{Algorithm \ref{ppo_train}}}
    \Statex \hspace{0.5em} \textbf{Test Step:}
      \State Randomly select $(b, b') \in (\mathbf{B}, \mathbf{B'})$
      \State Compute $g^c_{ij}$ as in Eq. (\ref{channel_gain}) and $\eta$ as in Eq. (\ref{sinr})
      \State $\xi \gets (q, \eta)$
      \For{$i \gets 1$ to \text{ITERATION}}
        \State $r, q, \varrho, \nabla, d, b \gets \text{Env}(\xi)$
        \State Compute $g^c_{ij}$ as in Eq. (\ref{channel_gain}) based on $b$
        \State Compute $\eta$ as in Eq. (\ref{sinr})
        \State Update $\xi \gets (q, \eta)$
      \EndFor
    \EndFor
    \State \Return Average reward $r$, queue length $q$, velocity resolution $\varrho$, data rate $\nabla$, and dropped packets $d$ over the evaluation phase.

    \end{algorithmic}
\end{algorithm}


\subsection{Values and Objectives Functions}
Almost all reinforcement learning algorithms rely on estimating value functions. These functions assess how beneficial it is for an agent to be in a specific state or to take a particular action in a given state, by predicting future rewards. The concept of "benefit" is fundamentally based on the expected return, which represents the anticipated future rewards. Since the future rewards depend on the agent's actions, value functions are determined with reference to specific policies.\\
\indent A policy $\pi$ at a given time slot $s$, is a function that maps each state, $\xi_s$, and action, $a_s$, by the probability of taking action $a_s$ when in state $\xi_s$. The value of a state $\xi_s$ under a policy $\pi$, represented as $V_\pi(\xi_s)$, is the expected return when starting from state $\xi_s$ and following the policy $\pi$ thereafter. This value can be defined in the aforementioned \gls{MDP} system  as follows \cite{9261348, chu2022ai}:
\begin{equation}
V_\pi(\xi_s):=\mathbb{E}_\pi\left[C(s) \mid \xi(s)=\xi_s\right].
\end{equation}
Here, \(\mathbb{E}_{\pi}\) represents the expected value when the agent follows policy \(\pi\), and \(s\) refers to any time step. The function \(V_\pi(\xi_s)\) is referred to as the state-value function for policy \(\pi\). \\
\begin{algorithm}
\caption{Agent Training}\label{ppo_train}
\begin{algorithmic}[1]
    \Require{\text {Initial policy parameters $\theta_0$ and initial value function} \text {parameters $\theta''_0$}} 
    \For {\text { for } k = 0,1,2, \ldots \text { do }}
        \State {Collect set of trajectories $\mathcal{D}_k=\left\{\tau^{\mathrm{tra}}_i\right\}$ by $\pi_k=\pi\left(\theta_k\right)$ $\mathrm{~~~~~}$in the environment.}

        \State \text {Compute the weighted cumulative rewards   $\hat{C}(s)$}
        \State Update $\theta_k$ using Eq. (\ref{ppo_objective_func}): 
         $\theta_{k+1}=\arg \max _\theta L(\theta)$ $\mathrm{~~~~~}$using stochastic gradient ascent with Adam. 
        \State Update $\theta''_k$ by:$\theta''_{k+1}=\arg \min _{\theta''} \mathbb{E}_{\mathcal{D}_k}\left(V_\phi\left(s_t\right)-\hat{R}_t\right)^2$ using gradient descent algorithm.
    \EndFor%
\end{algorithmic}
\end{algorithm}
\indent In the same way, the value of taking action \(a_s\) in state \(\xi_s\) under policy \(\pi\), denoted as \(Q_{\pi}(a_s, \xi_s)\), is defined as the expected return starting from state \(\xi_s\), performing action \(a_s\), and then following policy \(\pi\) for subsequent actions. Based on the aforementioned \gls{MDP} system, it can be computed as follows \cite{9261348, chu2022ai}: 
\begin{equation}
Q_\pi(\xi_s, a_s):=\mathbb{E}_\pi\left[C(s) \mid \xi(s)=\xi_s, a(s)=a_s\right].
\end{equation}
We refer to \(Q_{\pi}(a_s, \xi_s)\) as the action-value function for policy \(\pi\).\\
\indent In addition to the value function and the action-value function, there is a third function that measures how good an action is compared to other actions on average. This concept is known as the advantage function. The advantage function \(A_{\pi}(\xi_s, a_s)\) for a policy \(\pi\) quantifies how much better it is to take a specific action \(a\) in state \(\xi\) at a given time slot $s$ compared to randomly choosing an action according to \(\pi(\cdot | \xi)\) following the policy $\pi$ thereafter. It is expressed as follows \cite{9261348, chu2022ai}:
\begin{equation}
A_\pi(\xi_s, a_s) := Q_\pi(\xi_s, a_s)-V_\pi(\xi_s).
\end{equation}
\gls{PPO} is a policy optimization algorithm that aims at finding a policy with a maximum advantage. It combines ideas from both trust region policy optimization and clipped surrogate objective methods to ensure stable and efficient training. This is achieved by performing multiple epochs of minibatch updates using the collected samples. \gls{A2C} is an actor-critic algorithm that leverages the advantages of both policy gradients and value-based methods. It consists of two components: an actor, which selects actions based on the current policy, and a critic, which estimates the value function and provides feedback on the quality of the actor's actions.\\
\indent Both \gls{PPO} and \gls{A2C} are policy gradient methods that improve a parameterized control policy through gradient descent. Unlike traditional value function approximation methods, where policies are derived from a value function, policy gradient methods focus on directly optimizing a policy to maximize the expected return within a specified policy class. In this approach, given a parameterized policy \(\pi_{\theta}\) with parameters \(\theta\) and a state \(s\), the algorithm produces the probability distribution for selecting an action \(a\).\\
\indent Policy gradient methods optimize an objective or a loss function. They estimate the policy gradient and then use it in a stochastic gradient ascent algorithm. We consider \gls{A2C} case where the objective function is expressed as \cite{NIPS1999_464d828b,9350833}: 
\begin{equation}\label{loss_a2c}
L^{A 2 C}(\theta)=\hat{\mathbb{E}}_s\left[\log \pi_\theta\left(a_s \mid \xi_s\right) \hat{A}_s\right],
\end{equation}
with $\pi_{\theta}$ is a stochastic policy parameterized by $\theta$, $A_s$ is an an estimator of the advantage function at timestep $s$.
By taking the gradient of the objective w.r.t. $\theta$, we get 
\begin{equation}
\nabla_\theta L^{A 2 C}(\theta)=\hat{\mathbb{E}}_t\left[\nabla_\theta \log \pi_\theta\left(a_s \mid \xi_s\right) \hat{A}_s\right].
\end{equation}
Let $\Xi_s(\theta)$ denote the probability ratio $\Xi_s(\theta) = \frac{\pi_\theta\left(a_s \mid \xi_s\right)}{\pi'_\theta\left(a_s \mid \xi_s\right)}$, with $\pi'_\theta$, the the previous or the old policy. The trust region algorithm maximizes the following objective function: 

\begin{equation}\label{loss}
    L(\theta)=\hat{\mathbb{E}}_s\left[\frac{\pi_\theta\left(a_s \mid \xi_s\right)}{\pi'_\theta\left(a_s \mid \xi_s\right)} \hat{A}_s\right] = \hat{\mathbb{E}}_s\left[\Xi_s(\theta) \hat{A}_s\right].
\end{equation}
Investigating Eq. (\ref{loss}), it results that a large change between consecutive policies can result in instabilities during the training of \gls{RL} algorithms. In addition, small changes in policy parameters $\theta$ can lead to significant changes in the resulting policy. Given that, gradient ascent cannot fully solve this problem and would impair the sample efficiency of the algorithm. Therefore, the loss function $L(\theta)$ in Eq. (\ref{loss}), can be modified to keep $\Xi_s(\theta)$ close to 1.\\
\indent In comparison to \gls{A2C}, \gls{PPO}-Clip uses a clipping method. It involves constraining the objective function within a range called the clipping range. The coefficient of the clipped objective function measures the difference between the current and previous policies. This mechanism restricts the extent of policy changes at each update step. Preventing excessive policy updates helps maintain stability in the learning process.

The objective function of \gls{PPO} encourages the agent to choose actions that lead to greater rewards. The \gls{PPO} objective function is defined as \cite{NIPS1999_464d828b,9350833}:
\begin{equation}  \label{ppo_objective_func}
L^{\mathrm{PPO}}_{\mathrm{clip}}(\theta)=\mathbb{E}\left[\min \left(\Xi_s(\theta) \hat{A}_s, \operatorname{clip}\left(\Xi_s(\theta), 1-\epsilon, 1+\epsilon\right) \hat{A}_s\right)\right]
\end{equation}
\begin{figure}[t!] 
    \centering
    \includegraphics[width=3.5in]{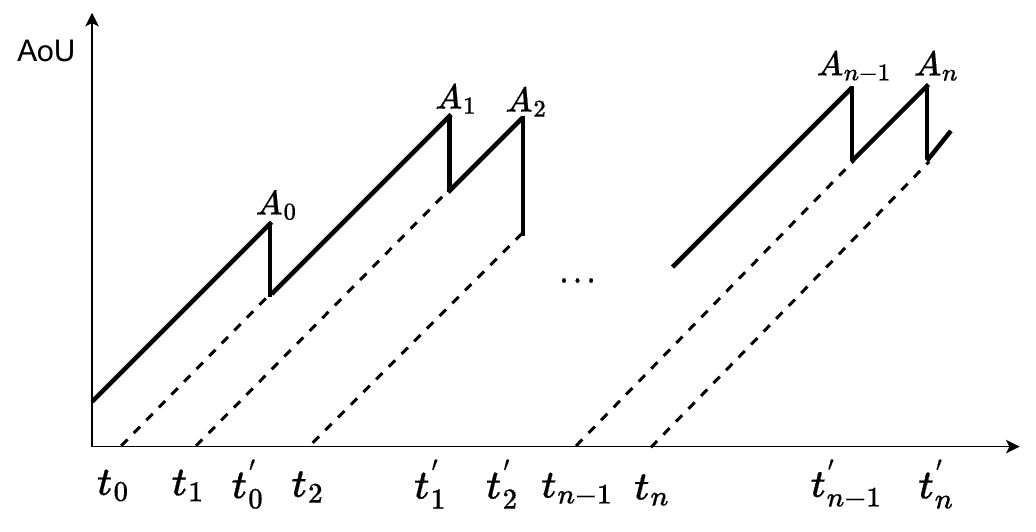}
    \caption{An age update process: updates are transmitted from a source at times $t_0$, $t_1$,$\dots$$t_n$ and are received at the receiver at times $t^{'}_0$, $t^{'}_1$,$\dots$$t^{'}_n$,
    and $A_0, A_1 \dots A_n$ are the corresponding age peaks.}
    \label{aou}
\end{figure}
Here, the \(\min\) operator selects the smaller value between the unclipped ratio \(\Xi_s(\theta) \hat{A}_s\) and the clipped ratio \(\operatorname{clip}\left(\Xi_s(\theta), 1-\epsilon, 1+\epsilon\right) \hat{A}_s\). \(\Xi_s(\theta)\)  measures how likely it is for the agent to take a particular action under the new policy compared to the old policy. It is then multiplied by the advantage function \(\hat{A}_s\). The advantage function \(\hat{A}_s\) evaluates the quality of an action relative to the average action within a given state. A higher value of \(\hat{A}_s\) indicates a greater expected improvement in outcome.

The function \(\text{clip}(\Xi_s(\theta), 1 - \epsilon, 1 + \epsilon)\) restricts \(\Xi_s(\theta)\) to the interval \([1 - \epsilon, 1 + \epsilon]\), where \(\epsilon\) is a small positive constant. This clipping mechanism ensures that the \gls{PPO} algorithm balances policy improvement with stability. The \gls{PPO} objective function guides the agent to make decisions that consider both the likelihood of actions under the new policy and their associated advantages.
 \begin{figure}[!t] 
    \centering
    \includegraphics[width=3.5in]{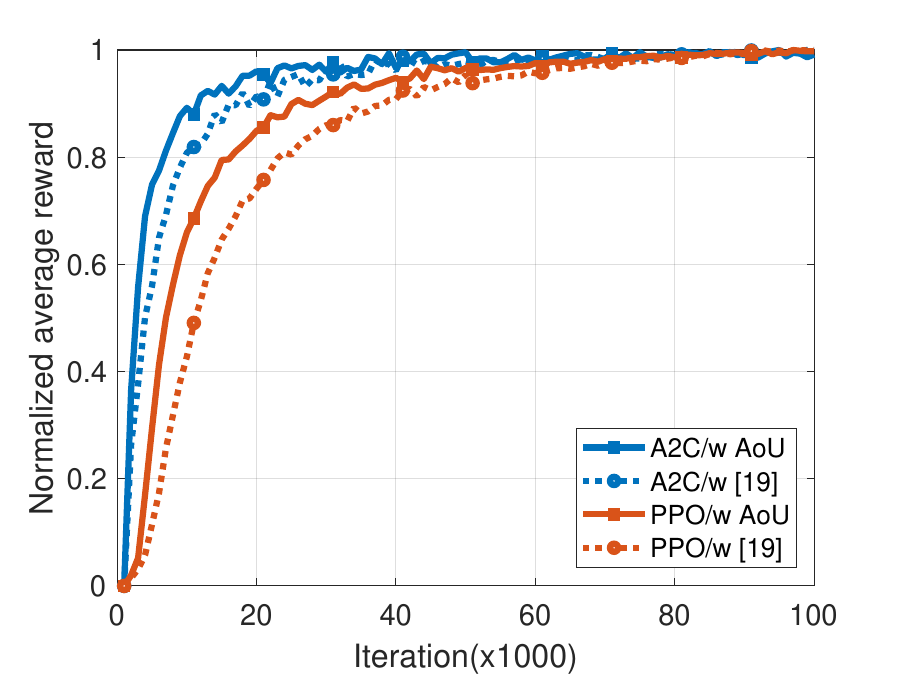}
    \caption{Normalized reward as function of the iterations.}
    \label{normalized average reward}
\end{figure}

\subsection{Age of Update-based Reward Modeling}
\indent The management of packets in the buffer presents several challenges, particularly regarding timeliness and effective decision-making. In real-time systems or environments with rapidly changing states, the freshness of information is crucial. For instance, the vehicular network may exchange time-critical information such as vehicle range and velocity to improve on-road safety. The vehicular network may be composed of several vehicles. For the onboard safety application of a given vehicle to perform well, the state of the other nodes of the network, i.e., the other cars and the surrounding objects, must be received at an acceptable effective data rate.\\
\indent Given the dynamism of the environment, the update frequency of the information must be high, which creates a potential congestion. Several works have proposed alternatives to detect the congestion. The authors in \textcolor{black}{\cite{9128822}} proposed a scheme for wireless communication between vehicles and roadside units, with congestion managed through packet rate control. They show that this approach reduces channel load and packet collisions by dropping packets. The work in \textcolor{black}{\cite{8398933}} proposes a congestion control scheme across transport, network, and MAC layers, using dual queue scheduling to prioritize transitory packets and dynamically update source sending rates to mitigate congestion. Several other techniques are suggested in \textcolor{black}{\cite{9933030, 9885572,10234061}}.
However, in this work, we introduce the \gls{AoU} metric to ensure that decisions are based on current and relevant data. By leveraging \gls{AoU}, a \gls{RL} agent can make more informed decisions based on up-to-date observations. This is especially important in \gls{V2V} communication where older information might be outdated and lead to sub-optimal actions.\\
\indent The average \gls{AoU} proposed in this work is derived from the age of information (AoI), a metric used to evaluate the freshness of information. It is defined as the duration that has passed since the generation of the most recent status update that was received as depicted in \mbox{Fig. \ref{aou}}. Specifically, if update $u$ was generated at time $e_u$ and delivered at time $v_u$, then AoI is calculated as follows is \cite{average_AoI1}: 
\begin{equation}\label{AoI}
    AoI(t) = \{t - \max\{e_u ~|~v_u \le t \}\}.
\end{equation}
This formula represents the time difference between the current time $t$ and the time of the latest successfully received update. The age increases linearly over time when no updates are received and is reset to a lower value upon receiving an update. Based on this definition and knowing that our \gls{MDP} is time slotted system, where $s$ a single slot time, $t= \{s, s+1, s+2, \dots s+N\} = \{s, 2s, 3s, \dots (N+1)s\}$.\\
\indent We propose the \gls{AoU} of a given packet as the time it spends from its arrival in the buffer $q$ to the end of its transmission. Therefore, from Eq. (\ref{AoI}), the \gls{AoU} of a given packet increments by one slot time if it is not received yet and is reset to 0 after its complete reception. Let us consider $t_q$ the queuing time, i.e.,  the time a given packet spends in the buffer before being propagated, and by $t_t$, the transmission delay and $t_p$, the propagation delay, the update time $t_a$ at the slot $s$ is expressed as: \\
\begin{equation}
    t_a(s) = t_q(s) + t_t(s) + t_p(s),
\end{equation}
with $t_p$ = $\frac{\bar{d}}{c_0}$ where $\bar{d}$ is the length of the \gls{V2V} link. The transmission delay $t_t$, which is negligible depends on the packet's size and the physical layer's effective data rate. Note that for every slot $s$, the delay $t_q$ is either increased to one slot time if the packet is still in the buffer or reset to zero once the packet is transmitted.\\
\indent We assume that at the beginning of the communication, the buffer is empty, the queue state and the \gls{AoU} are initialized to zero.  Moreover, two steps are taken to manage the buffer: (i) some packets arrive in the buffer following a Poisson distribution and (ii) depending on the action chosen by the agent, the number of packets to transmit is deduced. The buffer state is updated as in Eq. (\ref{queue_update}). Therefore, the age of the $i$th packet successfully received at the other end of the \gls{V2V} link is described as follows:
\begin{equation}\label{age_of_update}
     AoU_i(s)= \begin{cases} t_{a_i}(s-1)+1, &\text { packet $i$ not transmitted}\\
     0, & \text { otherwise }\end{cases}.
\end{equation}
Furthermore, the average \gls{AoU} of the whole buffer is derived as follows: 
\begin{equation}
    AoU_{q}(s) = \frac{\sum_{i=0}^{q(s)-1}AoU_i(s)}{q(s)+1}.
\end{equation}
Hence, the proposed \gls{RF} is defined as follows:
\begin{equation}\label{reward_fun2}
r(s) = r\left(\xi(s), a(s)\right)=-\left(\omega_1 AoU_{q}(s) + \omega_2 \varrho(s) + \omega_3 d(s)\right).
\end{equation}

\indent The existence of the optimal policy $\pi^*$ defined in the Eq. (\ref{optimal_policy}) assumes that $C(s)$ described in Eq. (\ref{cum_reward}) is always defined. As proof, in any given state $\xi$, the number of packets arriving in the buffer entirely follows a Poisson distribution. Furthermore, the environment is constantly varying, and subsequently, the \gls{CSI} also changes. This ensures that any state of the system can be reached starting from any initial state, making the \gls{MDP} irreducible. Hence, the average reward $C(s)$ is defined. The irreducibility of the \gls{MDP} ensures that the agent can explore the entire state space, which is crucial for learning an optimal policy. In addition, the technical analysis regarding the convergence of the actor-critic method, including \gls{PPO}, is extensively detailed in \cite{holzleitner2021convergence}.\\ 
\indent The procedural steps of our approach are summarized in Algorithm \ref{pseudo_algo}. 
The system optimizes key parameters such as the weighted reward, queue length, velocity resolution, data rate, and dropped packets. Algorithm \ref{pseudo_algo} describes iterative training and evaluation of the agent in a simulated environment that considers channel conditions and different blocking scenarios. 
The training and testing phases are integrated to ensure that optimal policies are learned by the model, which is then evaluated. Env() function defines the environment in which the agent operates. 
Given the current state $\xi$ comprising queue length $q$ and \gls{SINR} $\eta$, the system's next step is calculated based on the prediction of the agent.
During the training step,  the agent interacts with the environment, Env(), updates the internal policy as in Algorithm \ref{ppo_train}, and learns from the rewards received. This phase allows the model to make decisions that maximize the cumulative reward.
After training the model, the test step evaluates the performance of the model. The test involves running the agent through the same environment using the trained model. The performance is measured by assessing the average reward, queue length, velocity resolution, data rate, and dropped packets across the evaluation phase. 
 \begin{figure}[!t] 
    \centering
    \includegraphics[width=3.5in]{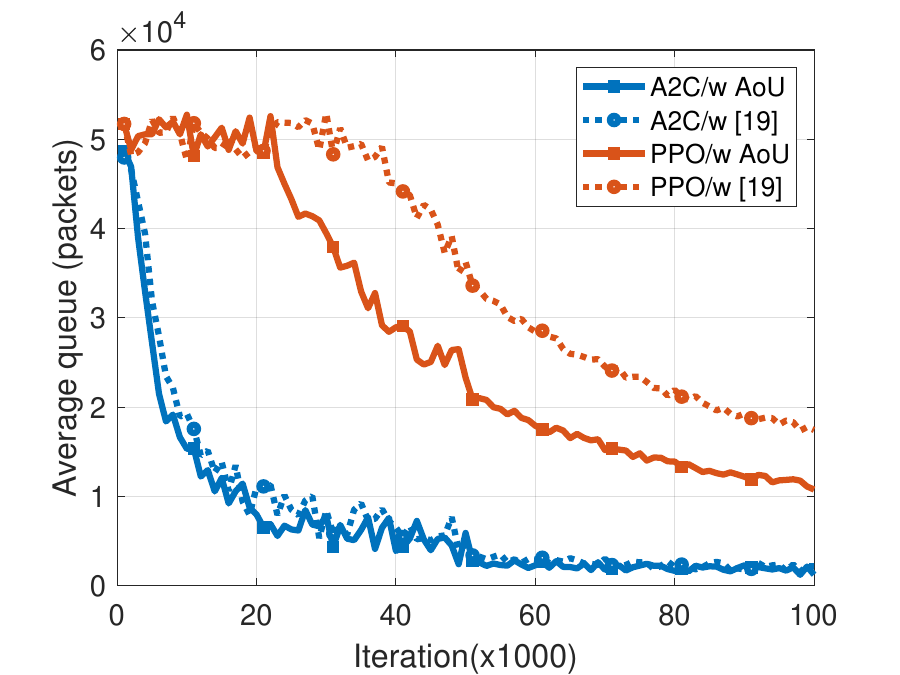}
    \caption{Average queue for a poor channel with $\lambda = 40,000$.}
    \label{p_queue}
\end{figure}

 \begin{figure}[!t] 
    \centering
    \includegraphics[width=3.5in]{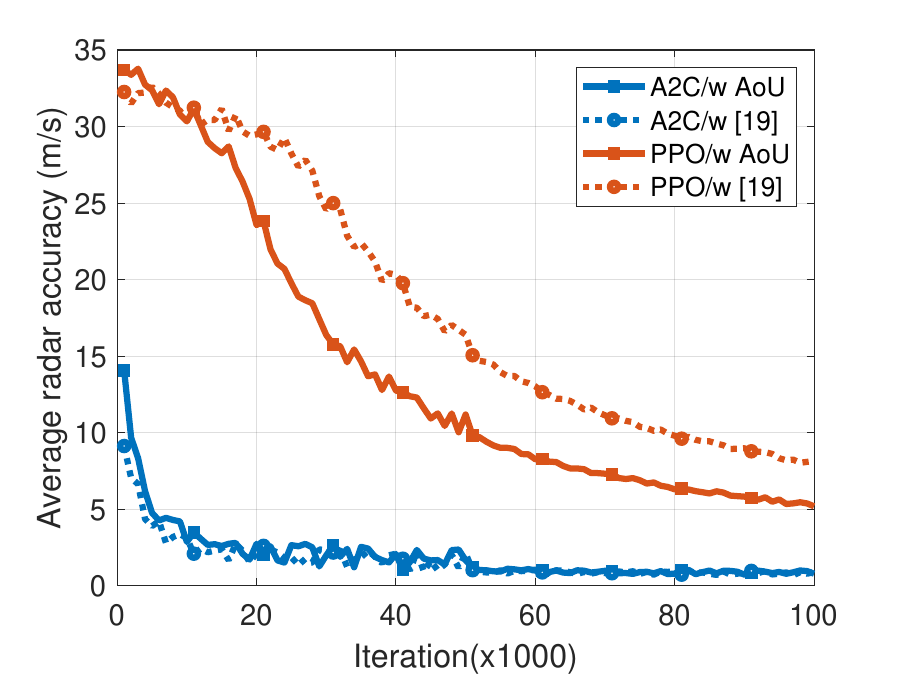}
    \caption{Average velocity accuracy for a poor channel with $\lambda = 40,000$.}
    \label{p_radacc}
\end{figure}

 \begin{figure}[!t] 
    \centering
    \includegraphics[width=3.5in]{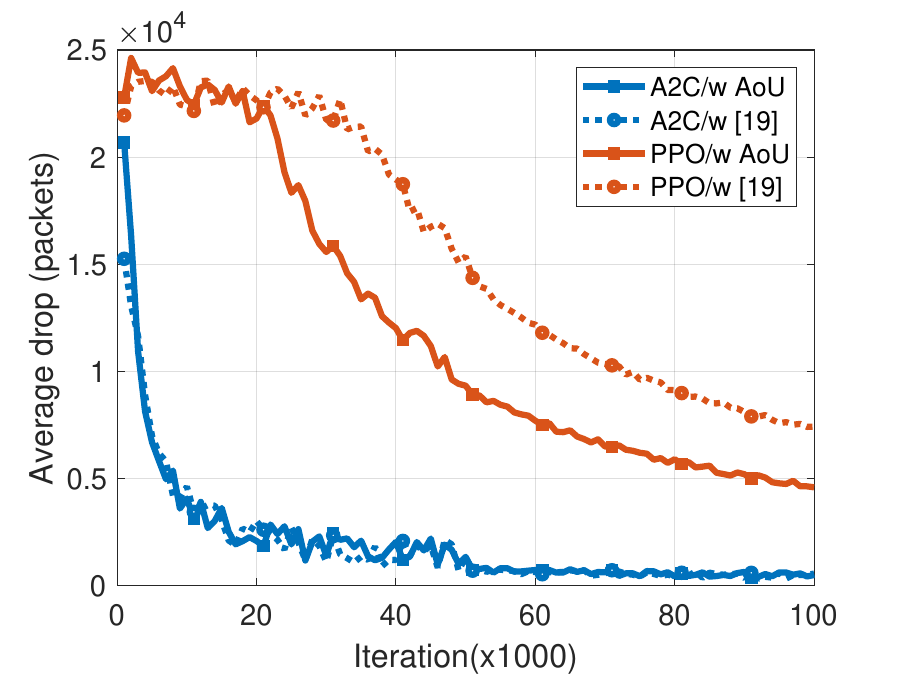}
    \caption{Average dropped packets for a poor channel with \mbox{$\lambda = 40,000$}.}
    \label{p_drop}
\end{figure}

 \begin{figure}[!t] 
\centering
\includegraphics[width=3.5in]{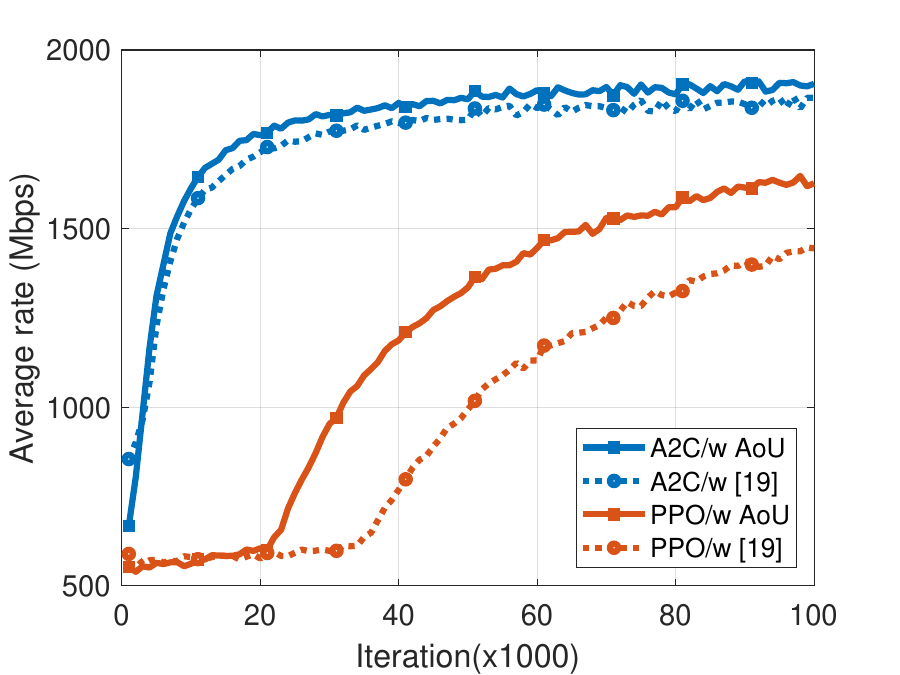}
\caption{Average effective data rate for a poor channel with $\lambda = 40,000$.}
\label{p_rate}
\end{figure}

\section{Simulation Results}
In this section, we assess the effectiveness of our approach in terms of communication and sensing. The simulations show the performance of \gls{PPO} and \gls{A2C} agents in optimizing \gls{V2V} links in a vehicular network.\\
\indent Through simulation results, we demonstrate that by incorporating \gls{AoU}, adaptive modulation, and \gls{SINR}, we achieve improvements in both communication and sensing.
To validate the effectiveness of the proposed reward function, we performed a comparison using \gls{PPO} and \gls{A2C} agents. The results, presented in Fig. \ref{normalized average reward}, clearly demonstrate that the inclusion of \gls{AoU} leads to a better cumulative reward, indicating a measurable improvement in both communication and sensing performance compared to the baseline reward function. Specifically, the AoU-based reward function better balances the trade-offs between communication efficiency and sensing accuracy by prioritizing timely updates, which are critical in dynamic environments. Additionally, we observe that the convergence using our scheme based on the \gls{AoU} is much faster compared to that of the approach in \cite{chu2022ai}.\\
\indent Improvements are also noticed in the other simulation results, specifically in terms of average queue state, average data rate, average discarded packet rate, and average velocity resolution in poor, normal, and strong channel conditions. These results are compared to the scheme proposed in \cite{chu2022ai},  which is the first that considers the dynamics of the environment throughout the learning process and does not consider \gls{AoU} and adaptive modulation.\\
\indent The simulation parameters are outlined in Table \ref{tab1}. In addition, \textit{poor}, \textit{normal}, and \textit{strong} channels are investigated with the packet error rate vector $\mathbf{B'}$  = [10\%,~1\%,~0.3\%] and blocker type $\mathbf{B} = ['LoS',~ '1V',~'2V',~'3V']$ stands for \gls{LoS}, one, two, and three vehicles between the intended transmitter and receiver, respectively. The probability vectors for \textit{poor}, \textit{normal}, and \textit{strong} links, associated with the type of blocker also called the obstruction probability are \mbox{$\mathbf{P}_p = [0.1, 0.1, 0.1, 0.7]$}, \mbox{$\mathbf{P}_n = [0.1, 0.7, 0.1, 0.1]$}, and \mbox{$\mathbf{P}_s = [0.7, 0.1, 0.1, 0.1]$}, respectively. The probability vectors for \textit{poor}, \textit{normal}, and \textit{strong} links, associated with the packets error rate are such that \mbox{$\mathbf{P'}_p = [0.8, 0.1, 0.1]$}, \mbox{$\mathbf{P'}_n = [0.1, 0.8, 0.1]$}, and \mbox{$\mathbf{P'}_s = [0.1, 0.1, 0.8]$}, respectively. Moreover, each packet is assumed to have a fixed size of \mbox{$\lvert \mathcal{P} \rvert$ = 4000} bytes, the maximum number of iterations is set at $10^5$ with 100 episodes, and the learning rate for PPO is fixed at 3x$10^{-4}$. The weight vector for balancing the immediate reward function is given by $\omega$ = [$10 ^{-5}$, 1, $10 ^{-5}$], i.e., \mbox{$\omega_1$ = $10 ^{-5}$}, \mbox{$\omega_2$ = 1}, \mbox{$\omega_3$ = $10 ^{-5}$}.\\
 \begin{figure}[!t] 
    \centering
    \includegraphics[width=3.5in]{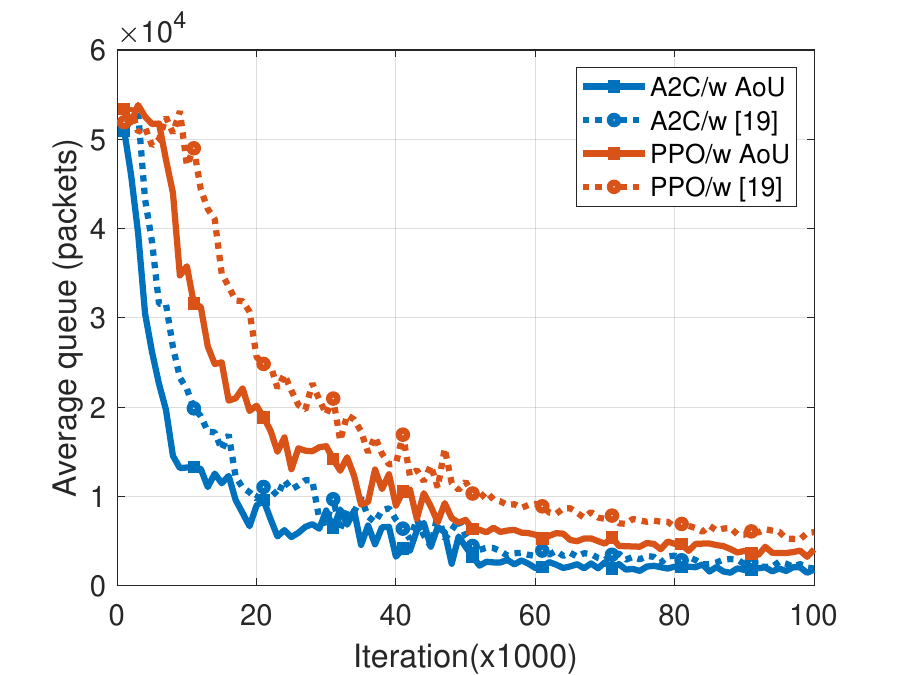}
    \caption{Average queue for a normal channel with \mbox{$\lambda = 120,000$}.}
      \label{n_queue}
\end{figure}
 \begin{figure}[!t] 
    \centering
    \includegraphics[width=3.5in]{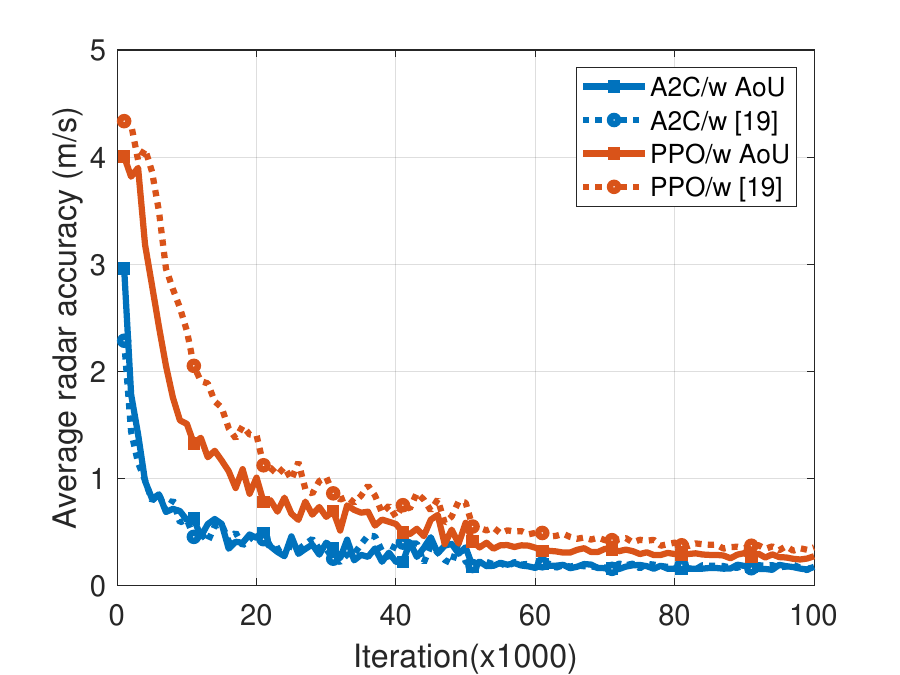}
    \caption{Average radar accuracy for a normal channel with $\lambda = 120,000$. }
      \label{n_radacc}
\end{figure}
 \begin{figure}[!t] 
    \centering
    \includegraphics[width=3.5in]{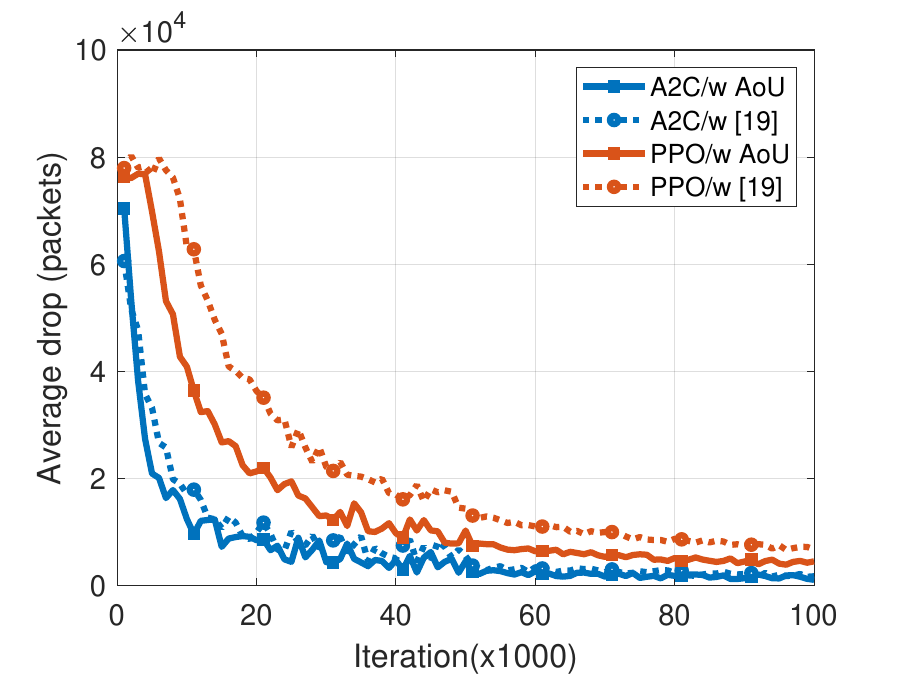}
    \caption{Average dropped packets for a normal channel with $\lambda = 120,000$. }
      \label{n_drop}
\end{figure}
 \begin{figure}[!t] 
    \centering
    \includegraphics[width=3.5in]{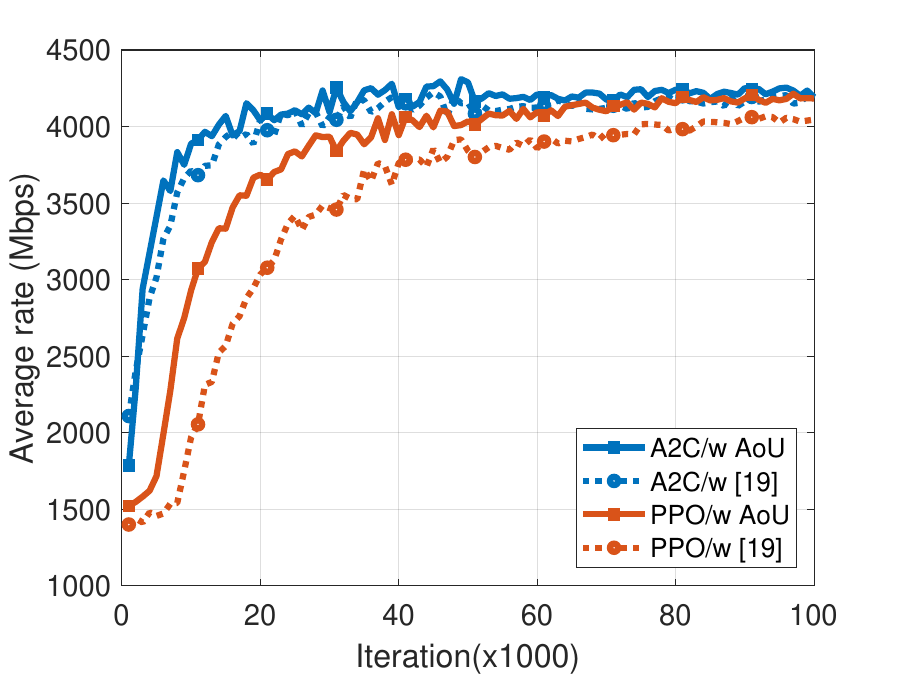}
    \caption{Average effective data rate for a normal channel with $\lambda = 120,000$. }
      \label{n_rate}
\end{figure}
 \begin{figure}[!t] 
    \centering
    \includegraphics[width=3.5in]{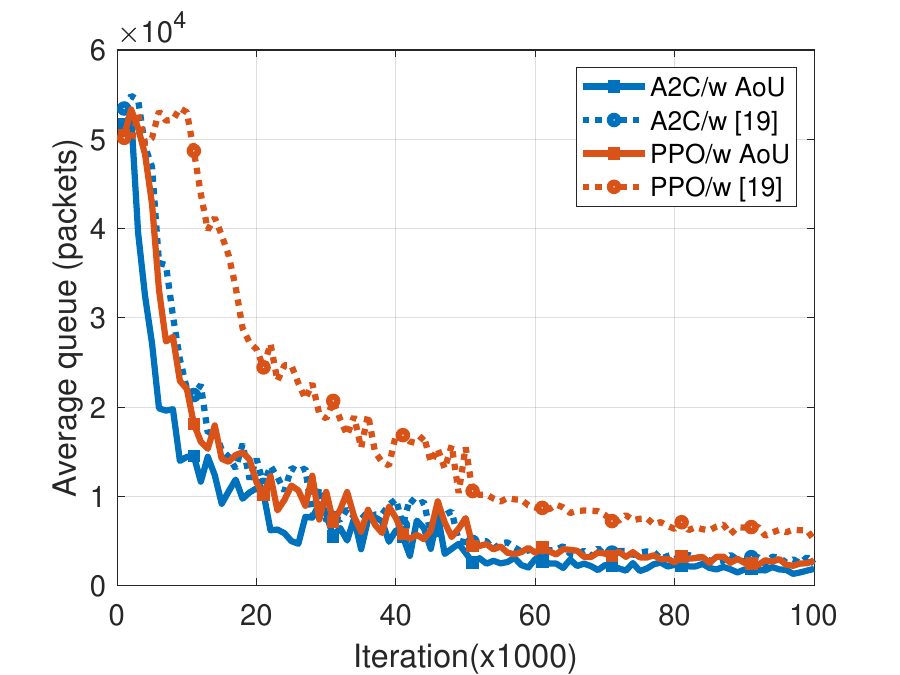}
   \caption{Average queue for a strong channel with \mbox{$\lambda = 180,000$}.}
     \label{s_queue}
\end{figure}
 \begin{figure}[!t] 
    \centering
    \includegraphics[width=3.5in]{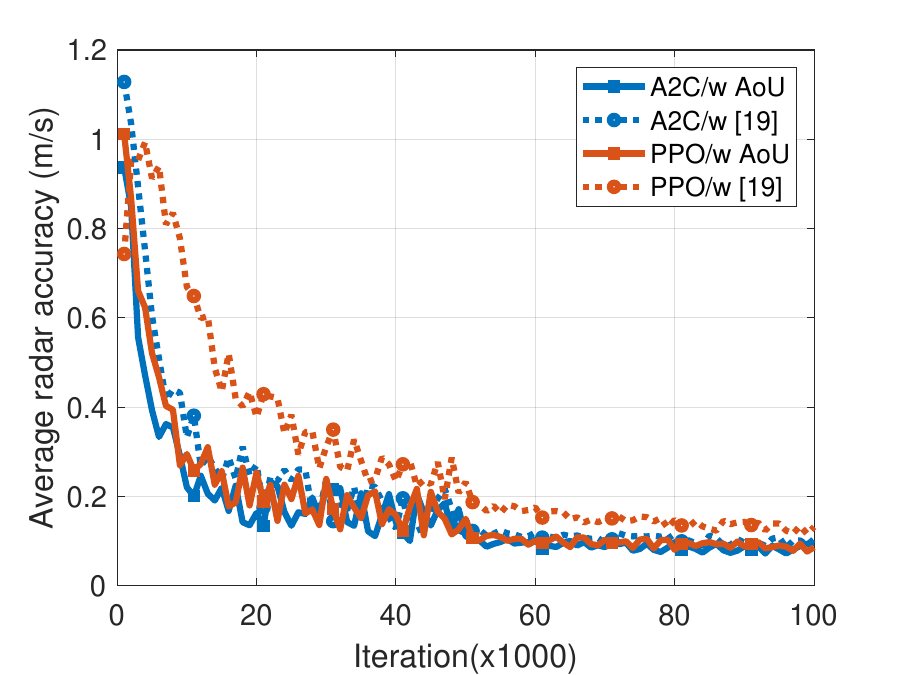}
   \caption{Average radar accuracy for a strong channel with $\lambda = 180,000$.}
     \label{s_radacc}
\end{figure}
 \begin{figure}[!t] 
    \centering
    \includegraphics[width=3.5in]{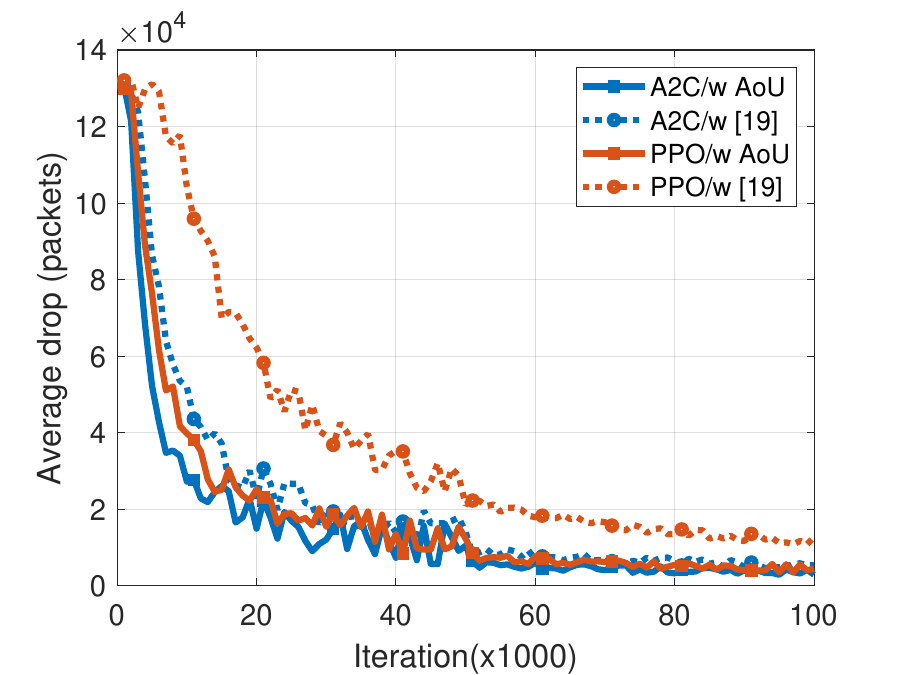}
   \caption{Average drop for a strong channel with $\lambda = 180,000$.}
     \label{s_drop}
\end{figure}
 \begin{figure}[!t] 
    \centering
    \includegraphics[width=3.5in]{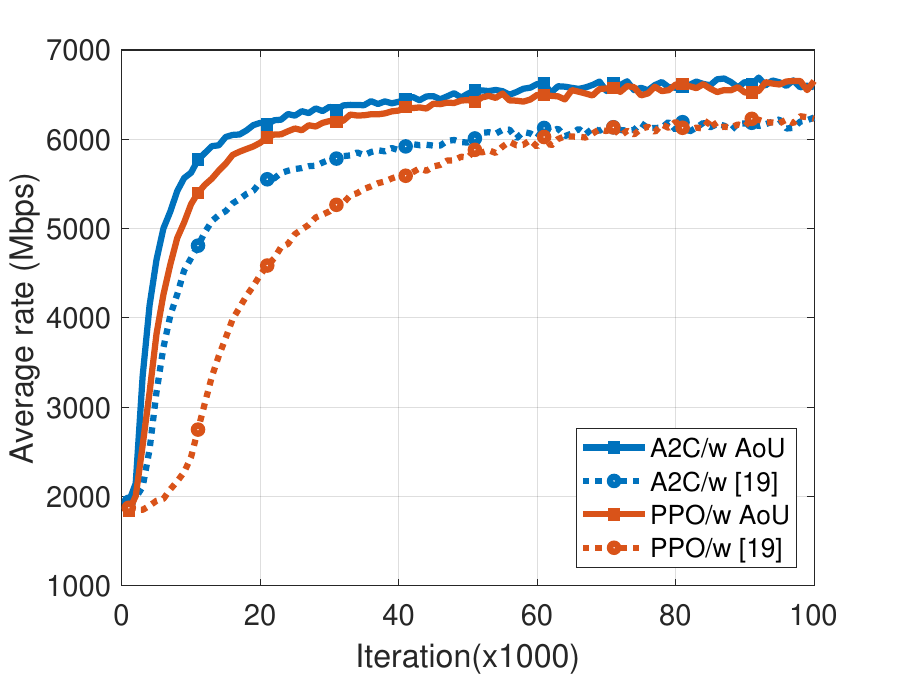}
   \caption{Average effective data rate for a strong channel with $\lambda = 180,000$.}
     \label{s_rate}
\end{figure}
\begin{table}[ht]
\caption{Simulation parameters}
\centering
\begin{tabular}{c c c }
\hline\hline
\textbf{Parameter} & \textbf{Symbol} & \textbf{Value} \\ [0.1ex] 
\hline
\vspace{0.08em}
Number of subcarriers&$N$&512\\
Maximum number of frames &$\bar{N}$&100\\
Modulations order &$\bar{m}$&[1, 2, 4, 6]\\
Carrier frequency &$f_c$&60 GHz \\
Wavelength&$\omega$&$5\times10^{-3}$ m \\
Bandwidth&$B_{\text{w}}$&2.16 GHz\\
Maximum range&$d_{max}$& 50 m\\
Maximum velocity&$v_{max}$& 50 m/s\\
Noise level&$N_0$& -174 dBm/Hz\\
Antenna gain &$g^a$& 1.5\\
Transmit power &$\rho_i$& 15 dBm\\
Interference  &$\mathbf{I}$&[24 , 24.5, 37, 37.5]~dB\\
Path loss exponent  &$\mathbf{\delta}$&[2.10, 1.22, 0.453, 0.240]\\
atmospheric attenuation  &$\mathbf{\beta}$&[75.1, 94.6, 126, 135]\\
Blocker type  &$\mathbf{B}$&['LOS', '1V','2V','3V']\\
Transmission slot time  &$s$&0.002\\
Pilote transmission time  &$T_p$&0.01*s\\
sector-level beamwidth  &$\psi$& $45^{\circ}$\\
half-power beamwidth   &$\varphi$& $10^{\circ}$\\
[1ex]
\hline
\end{tabular}
\label{tab1}
\end{table}
In Fig. \ref{normalized average reward}, we depict the normalized average reward as a function of the number of iterations. As shown in the graph, the return reward converges to a stable value. This indicates that after a certain number of iterations, the agent finds an optimal policy. The superior performance of the \gls{A2C} agent over the \gls{PPO} agent is explained by the fact that in our configuration, the environment changes very quickly due to the sensitivity of the \gls{V2V} link. Since the \gls{A2C} agent does not have the clipping mechanism as detailed in Eq. (\ref{loss_a2c}), it can sometimes make more significant updates to the policy, allowing it to adapt faster.\\
\indent In Fig. \ref{p_queue}, Fig. \ref{p_radacc}, Fig. \ref{p_drop} and  Fig. \ref{p_rate}, we show the average number of packets in the buffer, the average radar velocity resolution, the average number of dropped packets, and the average successfully delivered effective data rate, respectively, as a function of $\lambda$ in poor channel conditions with $\lambda = 40,000$. The \gls{PPO} agent shows a plateau phase particularly accentuated in Fig. \ref{p_queue}, \ref{p_drop}, and  \ref{p_rate}, where the queue length, the dropped packets, and the effective data rate stagnate first before they start converging. In poor channel conditions, the agent chooses actions that lead to a small transmit packet size. Consequently, the queue gets rapidly full by the incoming packets, which increases the packet drop probability and creates congestion that reduces the effective data rate. The \gls{A2C} demonstrates a notably better convergence, ultimately achieving a velocity resolution of 0.71 m/s and a effective data rate of 1.88 Gbps.\\
\indent In Fig. \ref{n_queue}, Fig. \ref{n_radacc}, Fig. \ref{n_drop} and Fig. \ref{n_rate}, we show the average number of packets in the buffer, the average radar velocity resolution, the average number of dropped packets, and the average successfully delivered effective data rate, respectively, as a function of $\lambda$ in normal channel conditions with $\lambda = 120,000$. The first remark is that the plateau phase is drastically reduced despite the higher number of packets arriving in the buffer compared to the poor channel conditions. After that, it is clear that our scheme exhibits better performance, ultimately achieving a velocity resolution of 0.14 m/s, which improves the poor channel velocity resolution by 80.28\%.  In addition, the average effective data rate achieves 4.18 Gbps which aligns with the IEEE 802.11ad SC-PHY with 16-QAM modulation, in which configuration the ideal  effective data rate is up to 4.62 Gbps \cite{va2016millimeter}.\\
\indent In Fig. \ref{s_queue}, Fig. \ref{s_radacc}, Fig. \ref{s_drop} and Fig. \ref{s_rate}, we show the average number of packets in the buffer, the average radar velocity resolution, the average number of dropped packets, and the average successfully delivered effective data rate, respectively, as a function of $\lambda$ in strong channel conditions with $\lambda = 180,000$. Despite a large number of packets arriving in the buffer, the scheme demonstrates its robustness. The agents can efficiently take convenient actions to avoid congestion and the overall performance is significantly enhanced. Remarkably, we can achieve a velocity resolution of 0.064 m/s which improves to 91\% the velocity resolution of poor channels.  We also notice that the effective data rate of 6.58 Gbps can be reached. This approximates the theoretical value, which is around 6.75 Gbps for IEEE 802.11ad \gls{OFDM} PHY with 64-QAM modulation. \cite{va2016millimeter}. \\
\indent These results underscore the efficiency of our approach in optimizing the \gls{V2V} link, even in challenging high-$\lambda$ scenarios, making it a promising choice for enhancing wireless communication systems. By leveraging the \gls{AoU}, we have proven how it is a valuable asset in a dynamic environment. By introducing the \gls{AoU} into the decision-making process, the model is empowered to optimize the waveform structure effectively, striking a balance between communication and sensing requirements. Our approach yields superior performance compared to the existing scheme, as shown in the detailed analysis of the provided figures.
\section{Conclusion}
In this work, we introduced a novel immediate \gls{RF} using the \gls{AoU} to optimize a \gls{V2V} link in a dynamic environment. We improved the buffer management and drastically reduced the packet loss. In addition, our scheme outputs enhanced radar resolution and high delivery effective data rates that can reach the IEEE 802.11ad PHY effective data rates. The seamless usage of adaptive \gls{OFDM} ensures that the system maintains an optimal trade-off between data throughput and error resilience, resulting in superior communication performance. To ensure the robustness of the scheme, we designed three scenarios based on poor, normal, and strong channel conditions. Through simulations, we demonstrated that our scheme adapts perfectly to these three scenarios, consistently achieving better performance compared to the previous work. Overall, our findings highlight the potential of introducing the \gls{AoU} into the \gls{RF} and the benefits it brings to waveform optimization in a dynamic environment. In future work, an extensive approach could be exploiting meta-learning to provide generalized models that adapt to different environments.  The introduction of meta-learning as an extension opens up avenues for further improving the optimization process and making it more adaptable to various environments.
\balance
\bibliographystyle{IEEEtran}
\bibliography{biblio_traps_dynamics}
\end{document}